\documentclass[final,3p,times,twocolumn]{elsarticle}

\usepackage{amssymb}
\usepackage{nccmath}
\usepackage{epsfig}
\usepackage{amsthm}
\usepackage{amsmath}
\usepackage{multirow}
\usepackage{subcaption}
\usepackage{ulem}
\usepackage{dblfloatfix}

\usepackage[table,xcdraw]{xcolor}

\usepackage{lineno}

\journal{Astroparticle Physics}

\begin{document}

\begin{frontmatter}



\title{Search for ultra-high energy photons through preshower effect with gamma-ray telescopes: study of CTA-North efficiency}


\author[a]{Kevin Almeida Cheminant}
\ead{kevin.almeida-cheminant@ifj.edu.pl}
\author[a]{Dariusz G\'ora}
\author[a,c]{David E. Alvarez Castillo}
\author[b]{Aleksander \'Cwik\l{}a}
\author[a]{Niraj Dhital}
\author[d]{Alan R. Duffy}
\author[a]{Piotr Homola}
\author[a]{Konrad Kopa\'nski}
\author[e]{Marcin Kasztelan}
\author[f]{Peter Kovacs}
\author[g]{Marta Marek}
\author[h]{Alena Mozgova}
\author[a,c]{Vahab Nazari}
\author[i]{Micha\l{} Nied\'zwiecki}
\author[j]{Dominik Ostrog\'orski}
\author[k]{Karel Smolek}
\author[a]{Jaros\l{}aw Stasielak}
\author[a]{Oleksandr Sushchov}
\author[l]{Jilberto Zamora-Saa}

\address[a]{Institute of Nuclear Physics Polish Academy of Sciences, Radzikowskiego 152, 31-342 Krakow, Poland}
\address[b]{University of Technology, Warszawska 24, 31-155 KraKow, Poland}
\address[c]{Joint Institute for Nuclear Research, Dubna, 141980 Russia}
\address[d]{Centre for Astrophysics and Supercomputing, Swinburne University of Technology, Hawthorn, VIC 3122, Australia}
\address[e]{National Centre for Nuclear Physics, Andrzeja Soltana 7, 05-400 Otwock-Swierk, Poland}
\address[f]{Institute for Particle and Nuclear Physics, Wigner Research Centre for Physics, 1121 Budapest, Konkoly-Thege Miklos ut 29-33, Hungary}
\address[g]{Amateur astronomer}
\address[h]{Astronomical Observatory of Taras Shevchenko, National University of Kyiv, Kyiv, Ukraine}
\address[i]{Institute of Telecomputing, Faculty of Physics, Mathematics and Computer Science, University of Technology, Warszawska 24, 31-155 Krakow, Poland}
\address[j]{AGH University of Science and Technology, Adama Mickiewicza 30, 30-059 Krakow, Poland}
\address[k]{Institute of Experimental and Applied Physics, Czech Technical University in Prague, Husova 240/5, 110 00 Prague 1, Czech Republic}
\address[l]{Universidad Andres Bello, Departamento de Ciencias Fisicas, Facultad de Ciencias Exactas, Avenida Republica 498, Santiago, Chile}

\begin{abstract}
As ultra-high energy photons (EeV and beyond) propagate from their sources of production to Earth, radiation-matter interactions can occur, leading to an effective screening of the incident flux. In this energy domain, photons can undergo $e^{+}/e^{-}$ pair production when interacting with the surrounding geomagnetic field, which in turn can produce a cascade of electromagnetic particles called \textit{preshower}. Such cascade can initiate air showers in the Earth's atmosphere that gamma-ray telescopes, such as the next-generation gamma-ray observatory Cherenkov Telescope Array, can detect through Cherenkov emission. In this paper, we study the feasibility of detecting such phenomena using Monte-Carlo simulations of nearly horizontal air showers for the example of the La Palma site of the Cherenkov Telescope Array. We investigate the efficiency of multivariate analysis in correctly identifying preshower events initiated by 40 EeV photons and cosmic ray dominated background simulated in the energy range 10 TeV -- 10 EeV. The effective areas for such kind of events are also investigated and event rate predictions related to different ultra-high energy photons production models are presented. While the expected number of preshowers from diffuse emission of UHE photon for 30 hours of observation is estimated around $3.3\times10^{-5}$ based on the upper limits put by the Pierre Auger Observatory, this value is at the level of $2.7\times10^{-4}$ ($5.7\times 10^{-5}$) when considering the upper limits of the Pierre Auger Observatory (Telescope Array) on UHE photon point sources. However, UHE photon emission may undergo possible "boosting" due to gamma-ray burst, increasing the expected number of preshower events up to 0.17 and yielding a minimum required flux of $\sim 0.2$ $\mathrm{km^{-2}yr^{-1}}$ to obtain one preshower event, which is about a factor 10 higher than upper limits put by the Pierre Auger Observatory and Telescope Array (0.034 and 0.019 $\mathrm{km^{-2}yr^{-1}}$, respectively).

\end{abstract}


\end{frontmatter}

\newpage
\section{Introduction}

\begin{figure*}[ht]
  \centering
  \includegraphics[width=16cm]{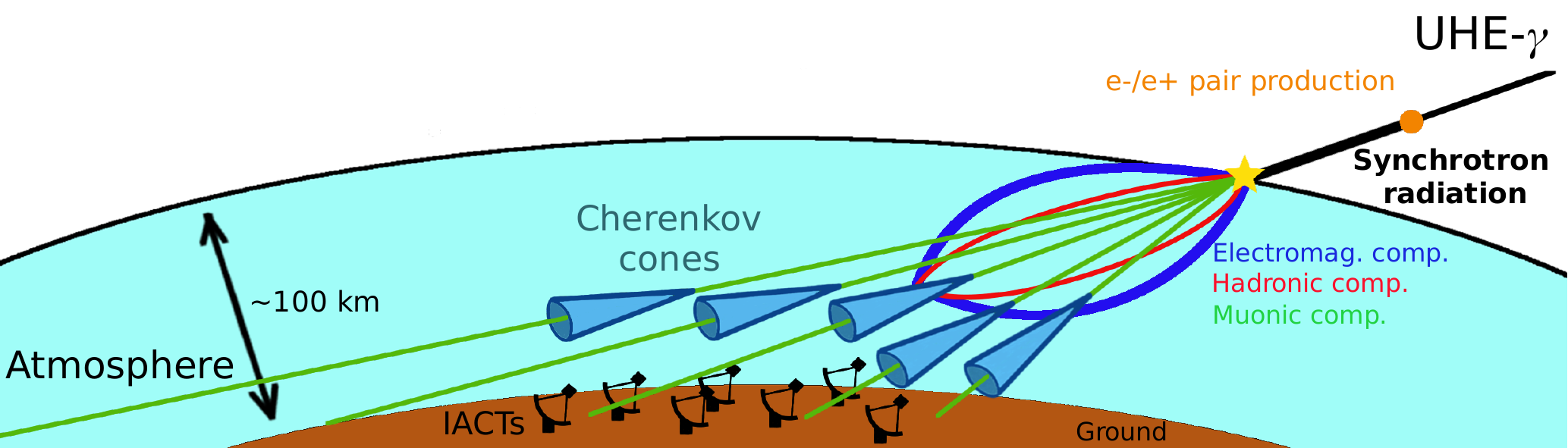}
  \caption{A ultra-high energy photon interacting with the transverse component of the geomagnetic field produces an $e^{+}/e^{-}$ pair $\sim$1000 km above sea level which emits bremsstrahlung photons. As such process can repeat itself for some of these photons, a collection of particles (mainly photons and a few $e^{+}$ and $e^{-}$) reaches the top of the atmosphere. Consequently, atmospheric air showers are produced and in the case of nearly horizontal showers, only the muonic component reaches the Imaging Atmospheric Cherenkov Telescopess (IACTs) on the ground, which detect the Cherenkov emission of this component.}
  \label{fig:sps}
\end{figure*}%

The observation of cosmic rays (CR) beyond the GZK cut-off (flux suppression due to the interaction of ultra-high energy cosmic rays (UHECRs) above $\sim5\times10^{10}$ GeV with the cosmic microwave background) \cite{greisen66,zatsepin66} constitutes a modern challenge for CR research and more generally, for astrophysics. Although the origin and the acceleration mechanisms of CRs in the MeV-TeV range seem to be well constrained \cite{bell78,tanimori98}, we have yet to understand how UHECRs above $10^{18}$ eV can reach such tremendous energies, and more broadly, where they are coming from. Cosmic rays in this energy range cannot be confined within our galactic disk by local magnetic fields, and no excess has been observed in the direction of the Milky Way. It is therefore believed that UHECRs are of extragalactic origins. Such assumption is supported by observations from Pierre Auger Observatory \cite{auger10} and Telescope Array \cite{abbasi14} which show anisotropies of UHECRs emission outside the galactic plane. However, these anisotropies do not seem to be clearly correlated with any known source powerful enough to generate these particles such as active galactic nuclei (AGN) or gamma-ray bursts. The recent detection of a 290 TeV neutrino by the ICECUBE collaboration \cite{icecube18}, whose arrival direction seems highly correlated to the position of the blazar TXS 0506+056 observed by FERMI-LAT and MAGIC \cite{multi18} has nonetheless provided evidence that AGN may in fact accelerate cosmic rays to the UHE domain. Another scenario for the production of UHECRs lies in so-called 'top-down' models. Among them, the decay of long-lived super-massive particles ($M_X > 10^{20}$ eV) \cite{chung98,kuzmin98,berezinsky00} may lead to a significant fraction of UHE photons \cite{berezinsky97} in the UHECRs flux, up to 50\%. UHE photons may also contribute to the CR flux as a product of the interaction of UHE protons with the cosmic microwave background and their observation could be a direct evidence of the GZK effect. However, in this scenario, the expected fraction of UHE photons is much lower, i.e. up to 10\% \cite{gelmini08}. Because the expected fraction of UHE photons in the UHECRs flux varies with the considered production models, detecting such photons is crucial for understanding the Universe in the UHE domain. 

The search for UHE photons therefore constitutes an important aspect of CR physics. Search or diffuse flux of UHE photons by the Pierre Auger Collaboration using nine-year of data set rules out any significant detection of such particles, on the assumption of a mixed composition for the CR background (50\% proton - 50\% iron) \cite{aab177}.

One possible explanation could lie in the extinction of the UHE photons flux as they propagate through space and interact with the geomagnetic field through the so-called \textit{preshower} effect \cite{mcbreen81}. In this scenario, a UHE photon produces an $e^{+}/e^{-}$ pair in the Earth's magnetic field (and to some extent, in any strong magnetic field such as the Sun's), which quickly loses energy via bremsstrahlung radiation. The resulting products of these interactions, most likely in the form of extensive electromagnetic showers above the Earth's atmosphere is mainly composed of lower energy photons with a small addition of $e^{+}/e^{-}$ pairs. 

The existing literature, although scarce, provides a solid background to any further investigation of such physical process. In \cite{stanev97}, it was shown that the number of photons radiated by the $e^{+}/e^{-}$ pair is increasing with the strength of the magnetic field experienced by the pair. The number of particles contained in air showers generated by the preshower was also investigated and was shown to be positively correlated to the energy of the primary photon. An interesting conclusion was that any observation of a dependence of the development of extensive air showers (EAS), observed at the highest energies, on the arrival direction could indicate that these EAS are in fact produced by UHE photons. In \cite{bednarek99}, the highest energy events recorded by various detectors such as Fly's Eye \cite{bird94} and AGASA \cite{hayashida94} were studied and, after considering them as UHE photons, the probability of them cascading in the geomagnetic field was calculated. This study showed that this probability was strongly dependent on the transverse magnetic field component and that very large fluctuations of the number of photons radiated by the $e^{+}/e^{-}$ pair were expected. Finally, the spatial distribution of particles at the top of the atmosphere was discussed in the case of a preshower occuring in the Sun's and Earth's vicinity. Although this distribution can extend over several kilometers when a UHE photon interacts with the Sun's magnetic field \cite{dhital19}, it can decrease to just a few centimeters if such a photon happens to pair produce via the geomagnetic field. The arrival time of these particles at the top of the atmosphere spans a very small time window and it is therefore likely that CR experiments on the ground would see these multiple particles interacting with the atmosphere as one single EAS. A comparison of EAS generated by unconverted photon was provided in \cite{bertou00}. It was shown that due to the Landau-Pomeranchuk-Migdal (LPM) effect \cite{lpm53}, EASs from unconverted photons would develop much deeper in the atmosphere, causing the maximum $X_{\mathrm{max}}$ at which the shower reaches its maximum to be much closer to the ground (that is to say that $X_{\mathrm{max}}$ is larger). In fact, \cite{risse07} showed that the preshower effect tends to lower the value of $X_{\mathrm{max}}$ as most photons in the preshower have energies lower than the threshold value required for the LPM effect to be significant. This causes preshower-induced EAS to have $X_{\mathrm{max}}$ values closer to the ones of nuclei-induced EAS. Because of much smaller numbers of muons in the former case, it was concluded that combining fluorescence techniques (to observe longitudinal profile of EAS) to detection from surface detectors (to observe the muon content) of the Pierre Auger Observatory should allow the identification of unconverted UHE photons and of EAS produced by the preshower effect. In \cite{homola07}, a strong directional dependence of the UHE photon first conversion was found by comparing simulations performed for the two sites of the Pierre Auger Observatory, North and South. It was also shown that the location itself had a great impact on the conversion probability by demonstrating that a larger transition region was obtained closer to the poles. Finally, \cite{albuquerque09} showed that the longitudinal profile of preshower-induced EAS might resemble the one of EAS produced by massive exotic hadrons, although some other features, like the muon content, would allow the discrimination between them.
\begin{figure*}[!t]
  \centering
  \begin{subfigure}{.5\textwidth}
    \centering
    \includegraphics[width=8cm]{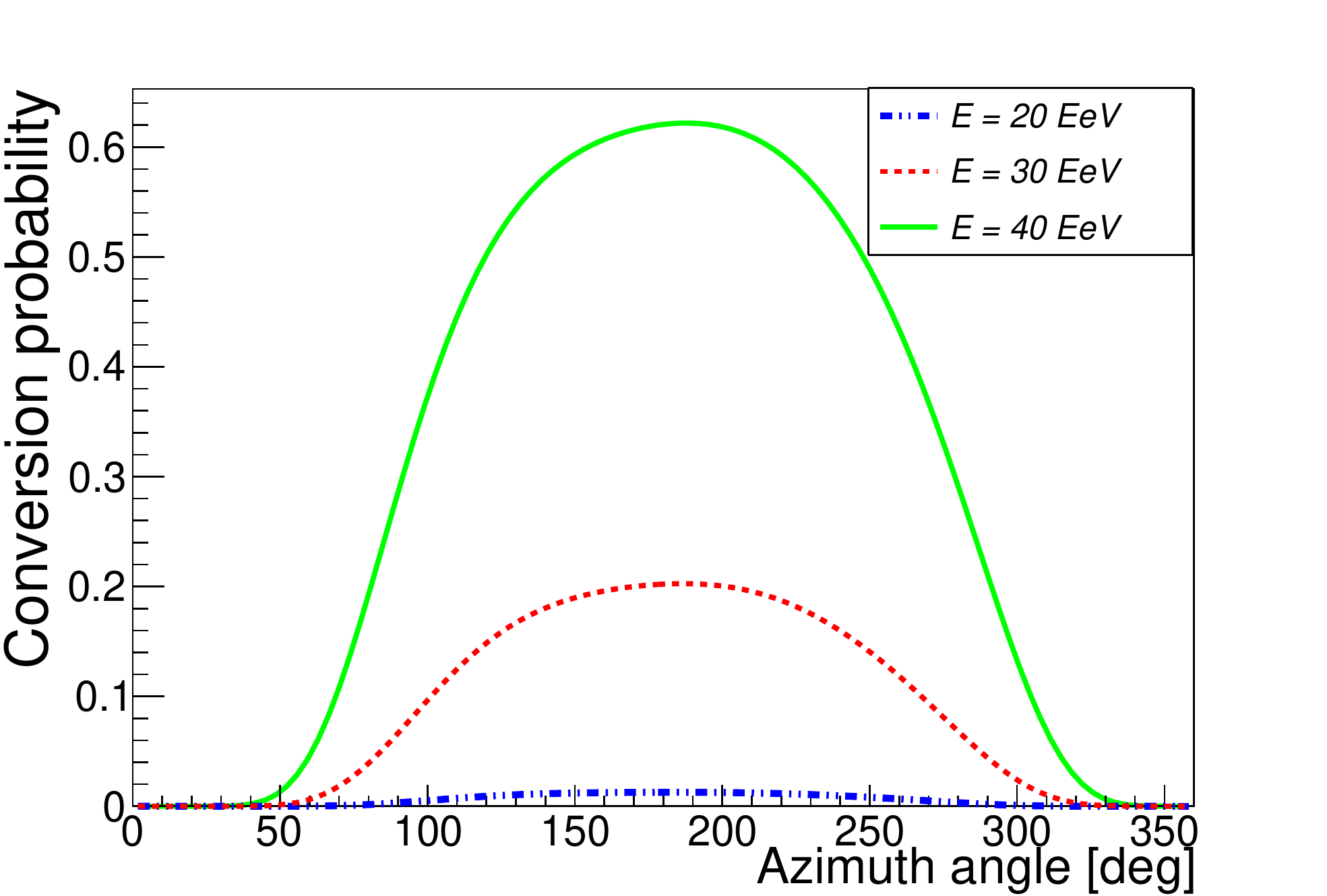}
    \label{fig:proba1}
  \end{subfigure}%
  \begin{subfigure}{.5\textwidth}
    \centering
     \includegraphics[width=8cm,height=5.5cm]{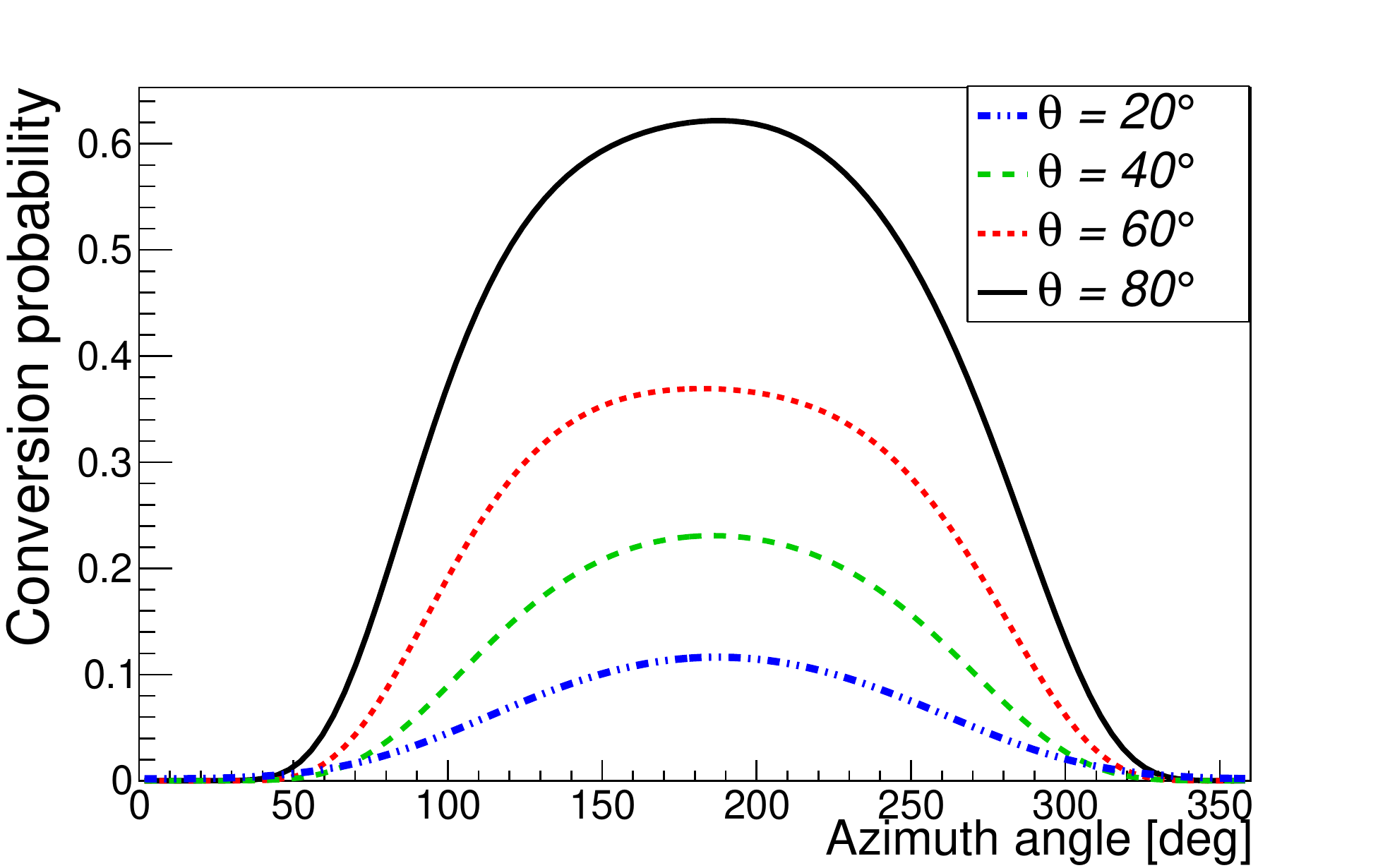}
    \label{fig:proba2}
  \end{subfigure}
    \caption{Probabilities of a UHE photon converting into an $e^{+}/e^{-}$ pair in the geomagnetic field. \textit{Left panel}: Conversion probabilities at zenith angle $\theta=80^{\circ}$ for different primary energies. \textit{Right panel}: Conversion probabilities for a 40 EeV primary photon at different zenith angles. Both plots are obtained for La Palma ($28^{\circ} 45' 43.2'' N, 17^{\circ} 53' 31.2'' W$) coordinates in CORSIKA frame of reference (azimuth angle, $\phi=0^{\circ}$ means that the particle comes from the geomagnetic South).}
  \label{fig:proba}
\end{figure*}
\begin{figure*}[t]
  \centering
  \begin{subfigure}{.5\textwidth}
    \centering
    \includegraphics[width=8cm]{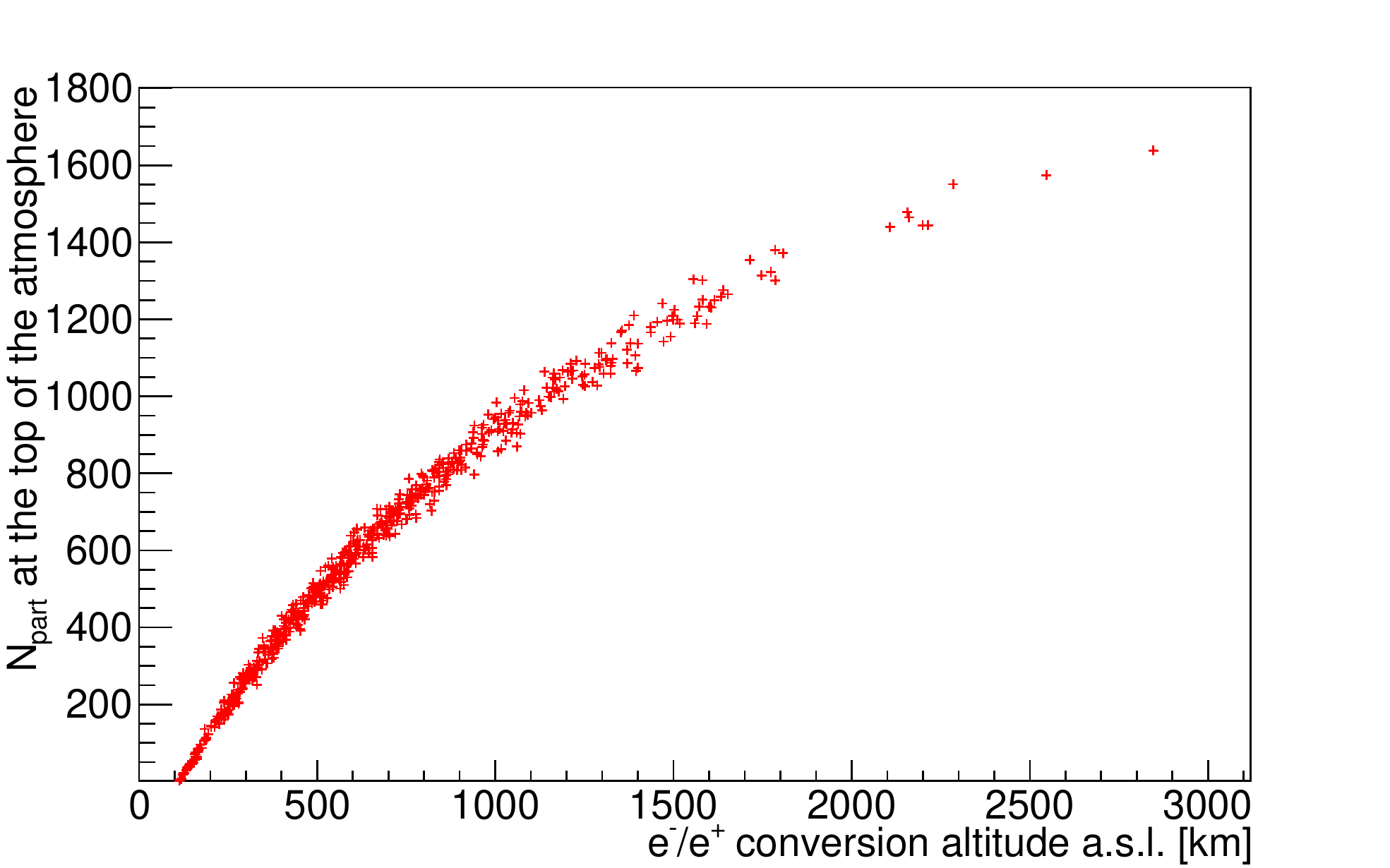}
    \label{fig:preshw1}
  \end{subfigure}%
  \begin{subfigure}{.5\textwidth}
    \centering
    \includegraphics[width=8cm]{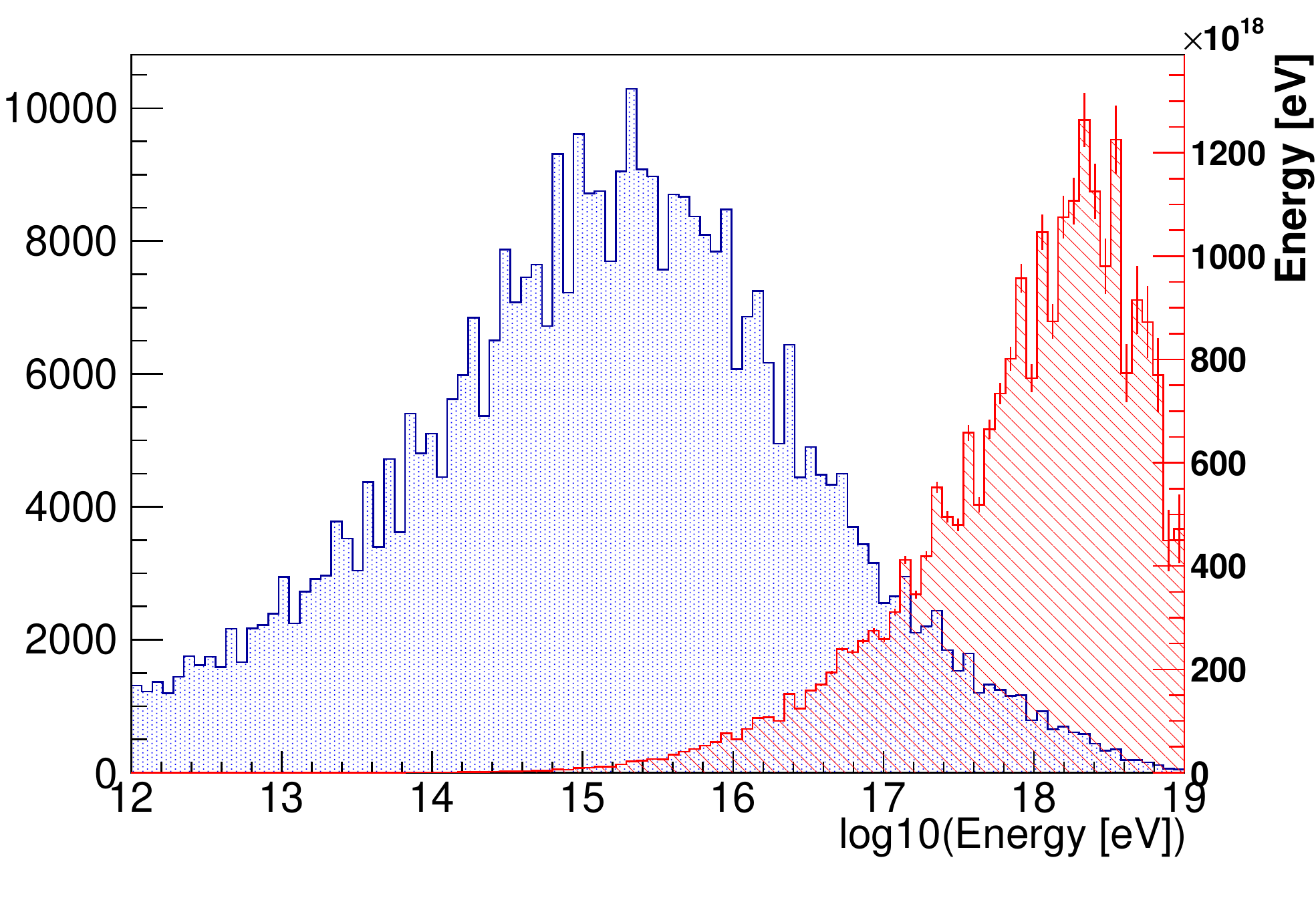}
    \label{fig:preshw2}
  \end{subfigure}
  \caption{\textit{Left panel}: Number of particles reaching the top of the atmosphere (at $\sim$100 km) as a function of the altitude at which a 40 EeV primary photon creates an $e^{+}/e^{-}$ pair. \textit{Right panel}: (Blue dotted histogram) Energy distribution of bremsstrahlung photons reaching the top of the atmosphere for 1000 simulations of UHE photon primary of 40 EeV, coming from direction defined by zenith and azimuth angles of $\theta=80^{\circ}$ and $\phi=180^{\circ}$, respectively. (Red hatched histogram) The same histogram weighted by energy (right Y-axis).}
  \label{fig:preshw}
\end{figure*}

It is clear that the extremely high energy regime that the preshower effect deals with naturally leads most research to look for such a phenomenon via experiments dedicated to the EeV domain. However, in this study, we argue that the preshower effect obtained via the geomagnetic field may also be studied with gamma-ray telescopes by using a non-standard observational approach introduced in \cite{neronov16}. In the TeV regime, the classic observation mode consists of pointing the telescope at fairly low zenith angles (closer to the vertical direction) in order for the cameras to collect enough Cherenkov light from air showers to obtain a good gamma/hadron separation. Some air showers might also be coming from directions closer to the horizon line but because of the much larger thickness of atmosphere implied by large zenith angles, only the muon component may survive to the ground. This component represents a good indicator of the nature of the particle that initiated the air shower and, because it has such a peculiar signature on the cameras of Cherenkov telescopes (muon rings), the muon component can be used to recover an acceptable gamma/hadron separation. Such a feature was in fact discussed in \cite{neronov16} and it was shown that the gamma/hadron separation could be recovered by collecting the Cherenkov light emitted by this component and by analyzing the images formed on the cameras. Such a method would have the potential to increase the sensitivity for the gamma-ray flux in different energy regime compared to various experiments like KASKADE-Grande and the Pierre Auger Observatory. Figure \ref{fig:sps} shows a schematic view of the processes studied in this paper: a nearly horizontal preshower produces a cascade of particles at the top of the atmosphere and the Cherenkov light emitted by the muonic component is detected on the ground by Imaging Atmospheric Cherenkov Telescopes (IACTs) while the other EAS components are mostly absorbed.

A preliminary study on the possibility to discriminate nearly horizontal air showers produced by protons, gammas and preshowers via the Cherenkov light detected by the cameras of the northern location of the next-generation Cherenkov Telescope Array \cite{cta11}(further referred as CTA-North) was presented in \cite{almeida17,almeida19}. In the present study, we extend the simulations of the CR background and the preshower effect by generating a CR energy spectrum from 10 TeV to 10 EeV and by investigating the influence of simulations parameters on the preshower/hadronic separation. The first section discusses the underlying physical characteristics of the simulated events along with the tools used to perform the simulations. In the second section, we briefly review the concept of boosted decision trees (BDT) as well as the variables used to discriminate preshowers from CR background. Finally, we present the results obtained from the multivariate analysis and calculate the aperture and expected number of preshower events, based on various UHE photons production models and upper limits set by different experiments.

\section{Simulations}

\subsection{Preshower effect}

The so-called preshower effect is simulated by the \textit{PRESHOWER} algorithm \cite{homola05} and results in a collection of particles (low energy gammas and $e^{+}/e^{-}$ pairs) reaching the top of the atmosphere. The first step in the simulation chain is the propagation of UHE photons in the geomagnetic field which is described by the International Geomagnetic Reference Field (IGRF) model \cite{igrf} and the calculation of the $e^{+}/e^{-}$ pair conversion probability. For a photon propagating over a distance $R$, the probability of converting to an $e^{+}/e^{-}$ pair is taken from \cite{erber66}:

\begin{equation}
 P_\mathrm{conv}(R) = 1 - \exp[ - \int\limits_{0}^{R} \alpha \left(\chi\right) dl], \label{eqn:P_conv}
\end{equation}
where $\alpha\left(\chi\right)$ is the photon attenuation coefficient and $\chi \equiv (1/2) (h \nu)/(m c^{2}) (B_{\bot})/(B_\mathrm{cr})$ is a dimensionless parameter depending on the perpendicular component of the geomagnetic field $B_{\bot}$ through which the photon is travelling and where $B_\mathrm{cr} \equiv \frac {m^{2} c^{3}}{e \hbar} = 4.414 \times 10^{13} \mathrm{\ G}$ is the natural quantum mechanical measure of magnetic field strength. Figure \ref{fig:proba} (left) shows conversion probabilities obtained for various energies. As expected, for a given arrival direction (fixed zenith and azimuth angles), the probability increases with the energy and is null for photons with energy of a few EeVs. On the other hand, Figure \ref{fig:proba} (right) shows that the conversion probability also increases with the zenith angle. Such behavior can be explained by the fact that the geomagnetic field increases as the altitude decreases. Therefore, for nearly horizontal arrival direction, the photon travels across a larger region of space where the geomagnetic field is the strongest. In both plots, one can notice that the probability is maximal for $\phi=180^{\circ}$ (UHE photon arriving from the geomagnetic North). This characteristic arises from the fact that the geomagnetic field gets stronger as we approach the geomagnetic axis. For the case of La Palma site, photons coming from a more northern direction tend to travel nearer this axis compared to the ones having a more southern trajectory. Consequently, they experience a stronger magnetic field and are more likely to create an $e^{+}/e^{-}$ pair.

If such conversion occurs, the $e^{+}/e^{-}$ pair travels along the direction of the primary photon and its trajectory is slightly deflected due to the geomagnetic field. The main energy loss process being bremsstrahlung radiation, the probability of an $e^{-}$ with energy $E$ to emit a photon between energy $h\nu$ and $h\nu+d(h\nu)$ over a distance $dx$ can be written as:

\begin{equation}
    P_\mathrm{brem}(B_{\bot},E,h\nu,dx) = dx\int\limits_{0}^{E} I(B_{\bot},E,h\nu)\frac{d(h\nu)}{h\nu}, \label{eqn:P_brem}
\end{equation}
where $B_{\bot}$ is the magnetic field component transverse to the primary photon direction and $I(B_{\bot},E,h\nu)$ is the spectral distribution of the radiated energy and is taken from \cite{erber66} (Eq. [2.5]). For bremsstrahlung photons of the highest energy, pair production can occur again and the whole process previously described may repeat itself. In the case of a 40 EeV primary photon, secondary photons do not create $e^{+}/e^{-}$ pairs. The actual number of bremsstrahlung photons produced strongly depends on the altitude at which the primary photon creates the first pair, as shown on the left plot of Figure \ref{fig:preshw}. Consequently, only the original $e^{+}/e^{-}$ pair and an ensemble of photons reach the top of the atmosphere. Figure \ref{fig:preshw} (right) shows the energy distribution of the secondary photons of the preshower down to $10^{12}$ eV obtained from the simulation of 1000 preshowers.

\begin{figure}[!b]
  \centering
  \includegraphics[width=5.5cm]{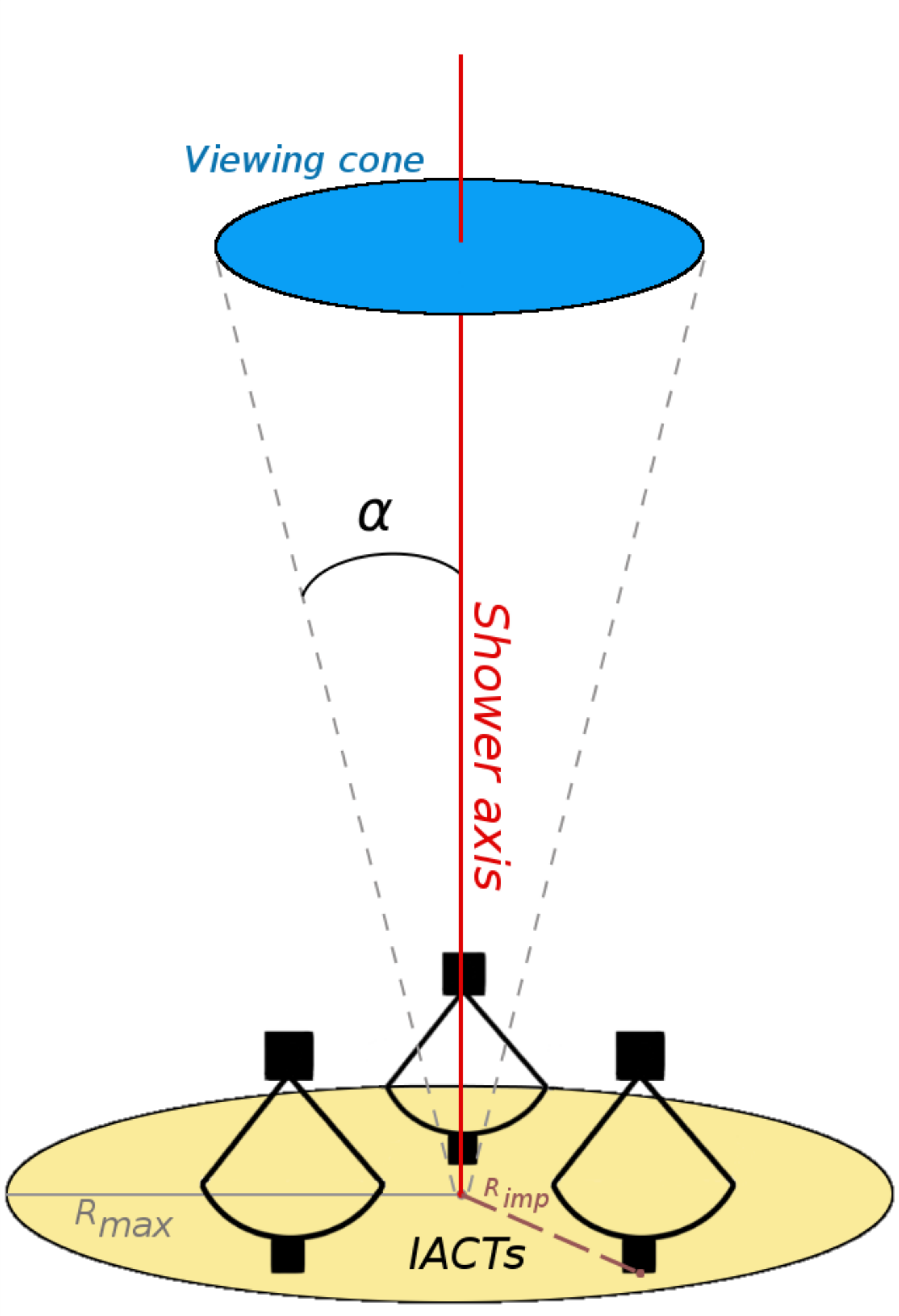}
  \caption{ Geometry and parameters of the CORSIKA simulations: The impact distance $R_{\mathrm{imp}}$ of the shower is randomly chosen in the interval [0;$R_{\mathrm{max}}$] and diffuse sources are generated with a randomly chosen angle within the cone defined by the opening angle $\alpha$.}
  \label{fig:corsika}
\end{figure}%

In order to get an idea of the spatial distribution of particles at the top of the atmosphere, one can calculate the linear displacement $\Delta x$ of the $e^{+}/e^{-}$ pair due to its deflection from the primary photon trajectory caused by the geomagnetic field. It can be approximated by:

\begin{equation}
    \Delta x \simeq \frac{L^2}{2R}, \label{eqn:P_displace}
\end{equation}
where $R$ is the radius of curvature of the $e^{+}/e^{-}$ trajectory and is approximately equal to $10^{13}$ km for a 20 EeV electron (the same for the positron which is deflected in the opposite direction than the electron due to its opposite charge) and for a typical transverse magnetic field of 0.1 G. With $L \simeq 1000$ km as the electron path length, we obtain $\Delta x \ll 1$ mm.

\begin{table*}[ht]
\centering
\resizebox{\textwidth}{!}{
\begin{tabular}{cllllllllllllll}
E [TeV]& \multicolumn{1}{c}{10} & \multicolumn{1}{c}{30} & \multicolumn{1}{c}{100} & \multicolumn{1}{c}{300} & \multicolumn{1}{c}{$10^{3}$} & \multicolumn{1}{c}{$3\times10^{3}$} & \multicolumn{1}{c}{$10^{4}$} & \multicolumn{1}{c}{$3\times10^{4}$} & \multicolumn{1}{c}{$10^{5}$} & \multicolumn{1}{c}{$3\times10^{5}$} & \multicolumn{1}{c}{$10^{6}$} & \multicolumn{1}{c}{$3\times10^{6}$} & \multicolumn{1}{c}{$10^{7}$}\\ \hline \hline
\multicolumn{1}{c}{$N_{\mathrm{trig}}$}&\multicolumn{1}{c}{121}&\multicolumn{1}{c}{446}&\multicolumn{1}{c}{865}&\multicolumn{1}{c}{1100}&\multicolumn{1}{c}{1219}&\multicolumn{1}{c}{1321}&\multicolumn{1}{c}{1382}&\multicolumn{1}{c}{1426}&\multicolumn{1}{c}{1437}&\multicolumn{1}{c}{1435}&\multicolumn{1}{c}{1435}&\multicolumn{1}{c}{1432}&\multicolumn{1}{c}{1436}\\  
\multicolumn{1}{c}{$dN_{\mathrm{CR,exp}}/dE\:[\mathrm{TeV}^{-1}]$}&\multicolumn{1}{c}{214238}&\multicolumn{1}{c}{38509}&\multicolumn{1}{c}{2725}&\multicolumn{1}{c}{169}&\multicolumn{1}{c}{7}&\multicolumn{1}{c}{0.36}&\multicolumn{1}{c}{0.01}&\multicolumn{1}{c}{$3\times10^{-4}$}&\multicolumn{1}{c}{$10^{-5}$}&\multicolumn{1}{c}{$4\times10^{-7}$}&\multicolumn{1}{c}{$10^{-8}$}&\multicolumn{1}{c}{$4\times10^{-10}$}&\multicolumn{1}{c}{$10^{-11}$}\\
\multicolumn{1}{c}{$N_{\mathrm{CR,exp}}(>E)$}&\multicolumn{1}{c}{1224780}&\multicolumn{1}{c}{660132}&\multicolumn{1}{c}{155649}&\multicolumn{1}{c}{28888}&\multicolumn{1}{c}{3829}&\multicolumn{1}{c}{540}&\multicolumn{1}{c}{51}&\multicolumn{1}{c}{6}&\multicolumn{1}{c}{0.5}&\multicolumn{1}{c}{0.06}&\multicolumn{1}{c}{0.005}&\multicolumn{1}{c}{$7\times10^{-4}$}&\multicolumn{1}{c}{$8\times10^{-5}$}\\
\hline \hline
\end{tabular}
}
\caption{\textit{Cosmic-ray background simulations} -- Number of CR events triggering the array (out of 1500 simulated showers), differential spectrum of expected CR events and number of events expected above an energy $E$, as a function of the simulated energy and for an observation time of 30 hours.}
\label{tab:trigger}
\end{table*}

In this study, we simulated 10000 showers for a point source and 10000 showers for a diffuse source in the case of $R_{\mathrm{max}}=1300$ m (see Section \ref{sec-corsika} for the definition of $R_{\mathrm{max}}$). In order to obtain the number of EAS simulated, these numbers must be multiplied by a conversion factor (defined by the probability for the UHE photon to produce an $e^{-}/e^{+}$ pair) of 0.67 in the considered direction of propagation as not all simulated UHE photons at 40 EeV convert.

\subsection{Atmospheric air showers} \label{sec-corsika}

The outcome of the PRESHOWER algorithm is then piped into CORSIKA 6.990 \cite{heck98}, a program that simulates the development of atmospheric air showers. The models selected for high and low energy hadronic interactions within the EAS are QGSJETII-03 \cite{ostapchenko06-1, ostapchenko06-2} and URQMD \cite{bass98, bleicher99}, respectively. Ultra-relativistic particles contained in the atmospheric cascade can travel faster than the speed of light in the air and produce Cherenkov emission. To account for this effect, the CERENKOV option was activated. In this work, we simulated the preshower effect for 2 different sets of parameters to investigate the preshower/CR background separation quality for different types of sources of UHE photons emission. In this endeavor, both CSCAT and VIEWCONE options were selected. The former enables randomization of the shower core on a disk (of radius $R_{\mathrm{max}}$) perpendicular to the shower axis, i.e. to the arrival direction defined by the zenith and azimuth angles of the primary particle. The latter defines a cone of apex angle $2\alpha$, with its apex pointing towards the detectors. Showers are generated within this cone, which allows simulations of both point ($\alpha = 0^{\circ}$) and diffuse ($\alpha > 0^{\circ}$) sources (see Figure \ref{fig:corsika}). In this study, the simulated showers are used only once.

\begin{figure}[!b]
  \centering
  \includegraphics[width=8cm]{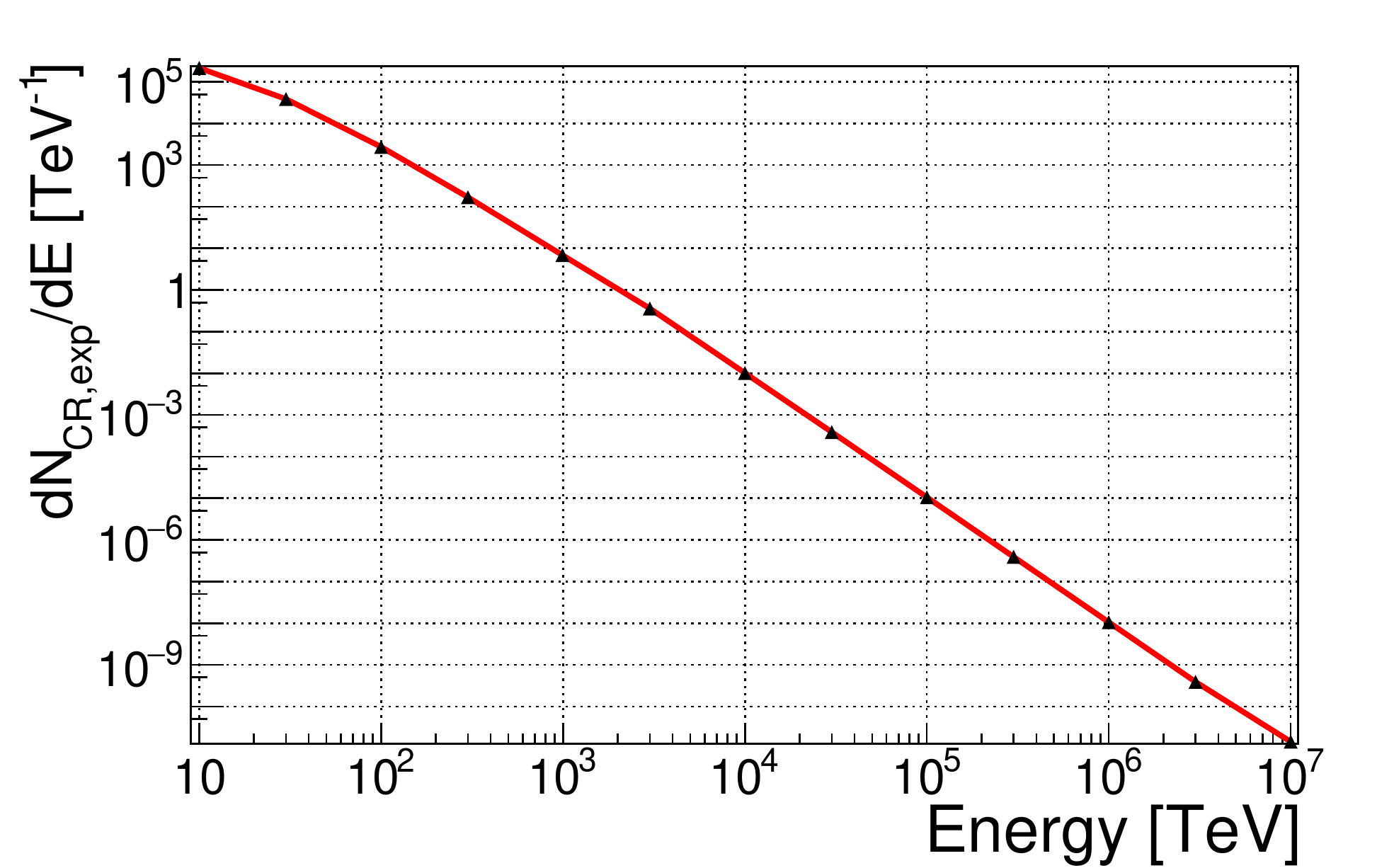}
  \caption{Number of expected CR background events for 30 hours of observation time with maximum impact distance of $R_{\mathrm{max}}=1300$ m obtained by convoluting the differential CR flux with the fraction of simulated air showers that triggers the array, the solid angle $\Omega$, the observation time $\Delta t$ and the simulated area $\pi R_{\mathrm{max}}^{2}$.}
  \label{fig:trigger}
\end{figure}%

Finally, because we consider inclined showers (EAS with large zenith angles), we choose to obtain their longitudinal profiles as a function of atmospheric depth measured along their axis, rather than as a function of the vertical atmospheric depth, by selecting the SLANT option. The CURVED EARTH option was also activated to properly account for the curvature of the atmosphere, which is especially important when EASs develop in the nearly horizontal direction. 

\begin{figure*}
  \begin{subfigure}{.5\textwidth}
    \centering
    \includegraphics[width=8cm]{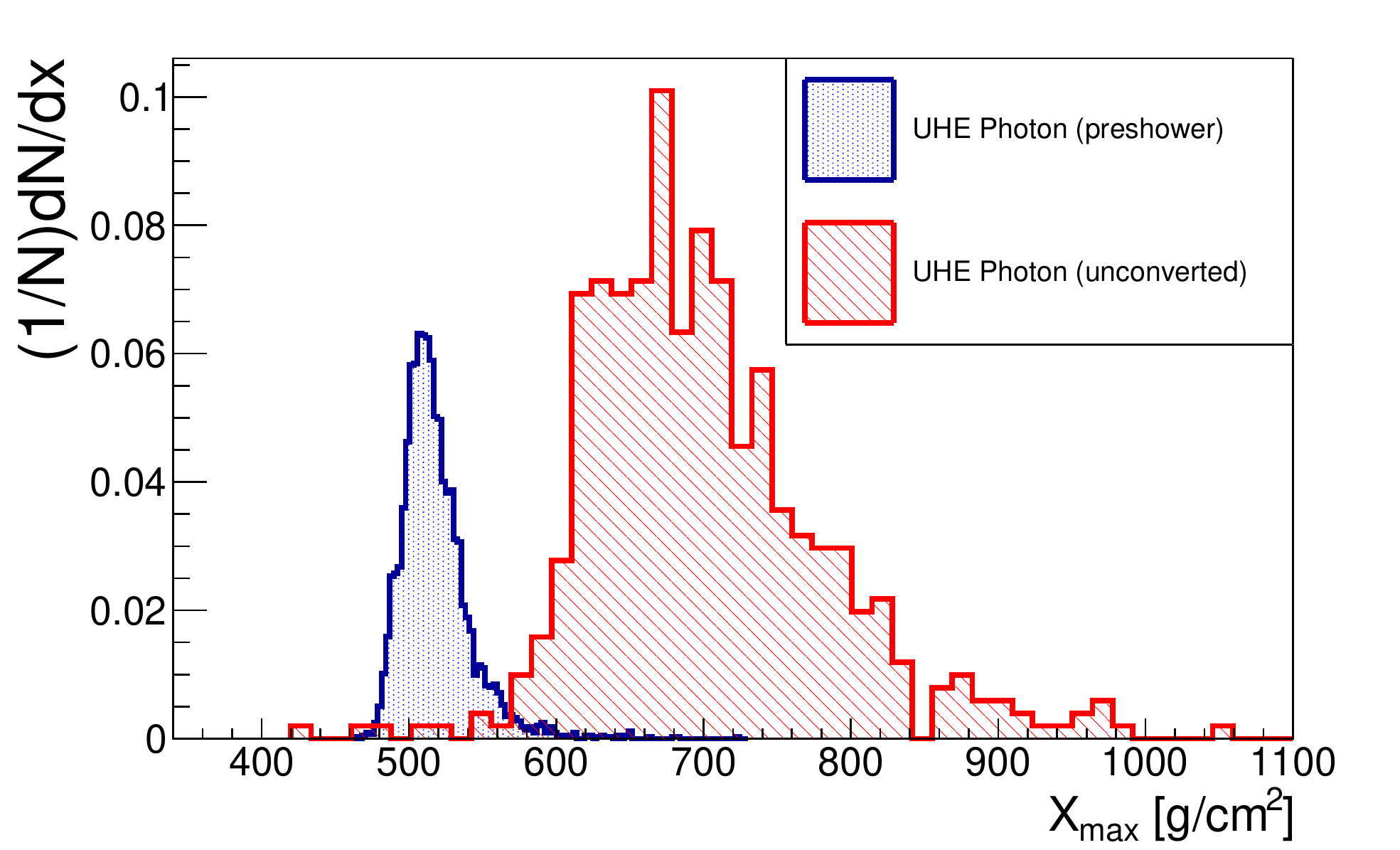}
  \end{subfigure}
  \begin{subfigure}{.5\textwidth}
    \centering
    \includegraphics[width=7cm,height=4.8cm]{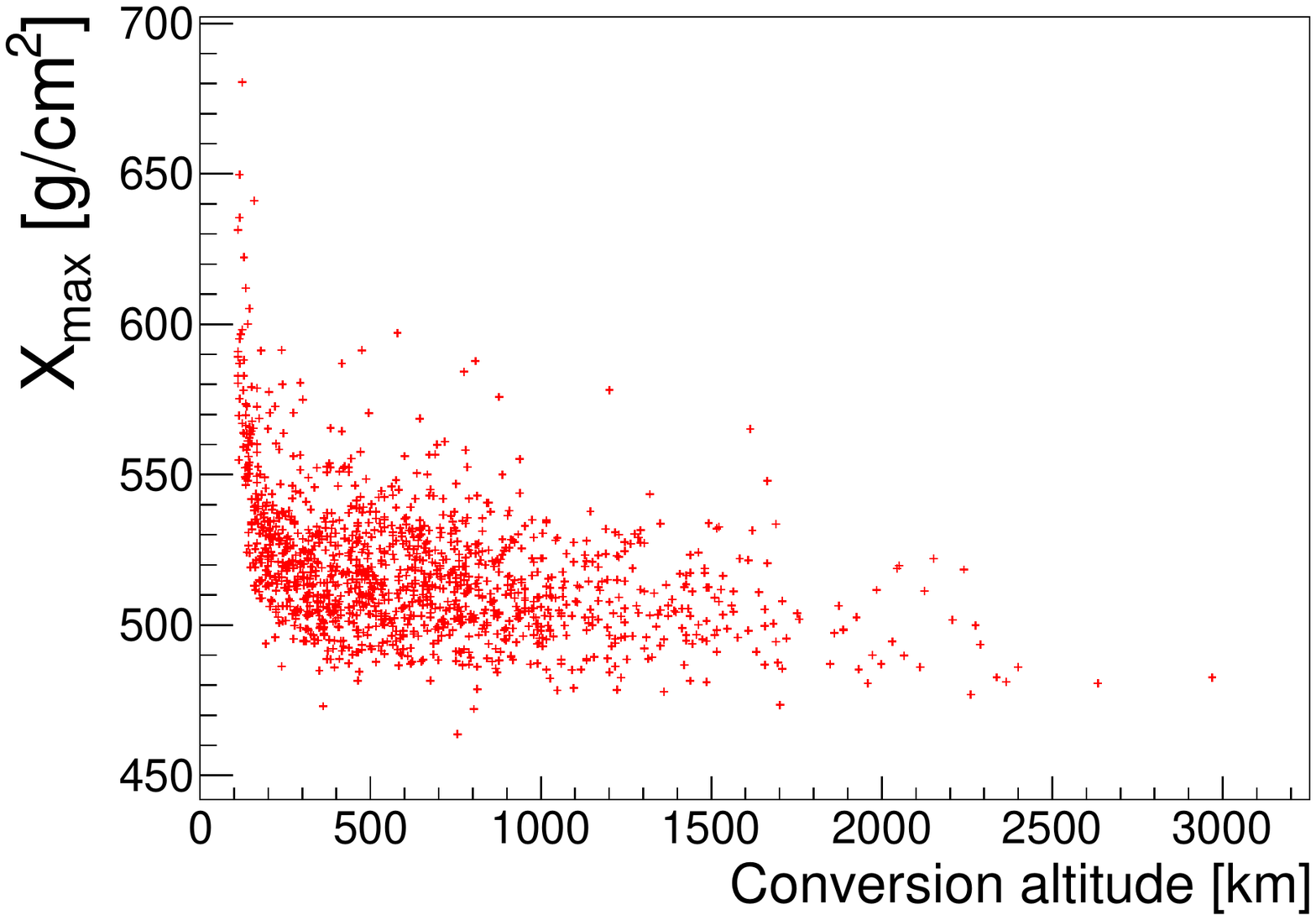}
  \end{subfigure}
  \caption{\textit{Left panel}: $X_{max}$ distribution of the nearly horizontal EAS initiated by 40 EeV primaries. \textit{Right panel}: Scatter plot of the $X_{max}$ distribution versus the altitude at which a 40 EeV UHE photon interacts with the geomagnetic field.}
  \label{fig:xmax}
\end{figure*}

\begin{figure*}[t]
  \begin{subfigure}{.5\textwidth}
    \centering
    \includegraphics[width=8cm]{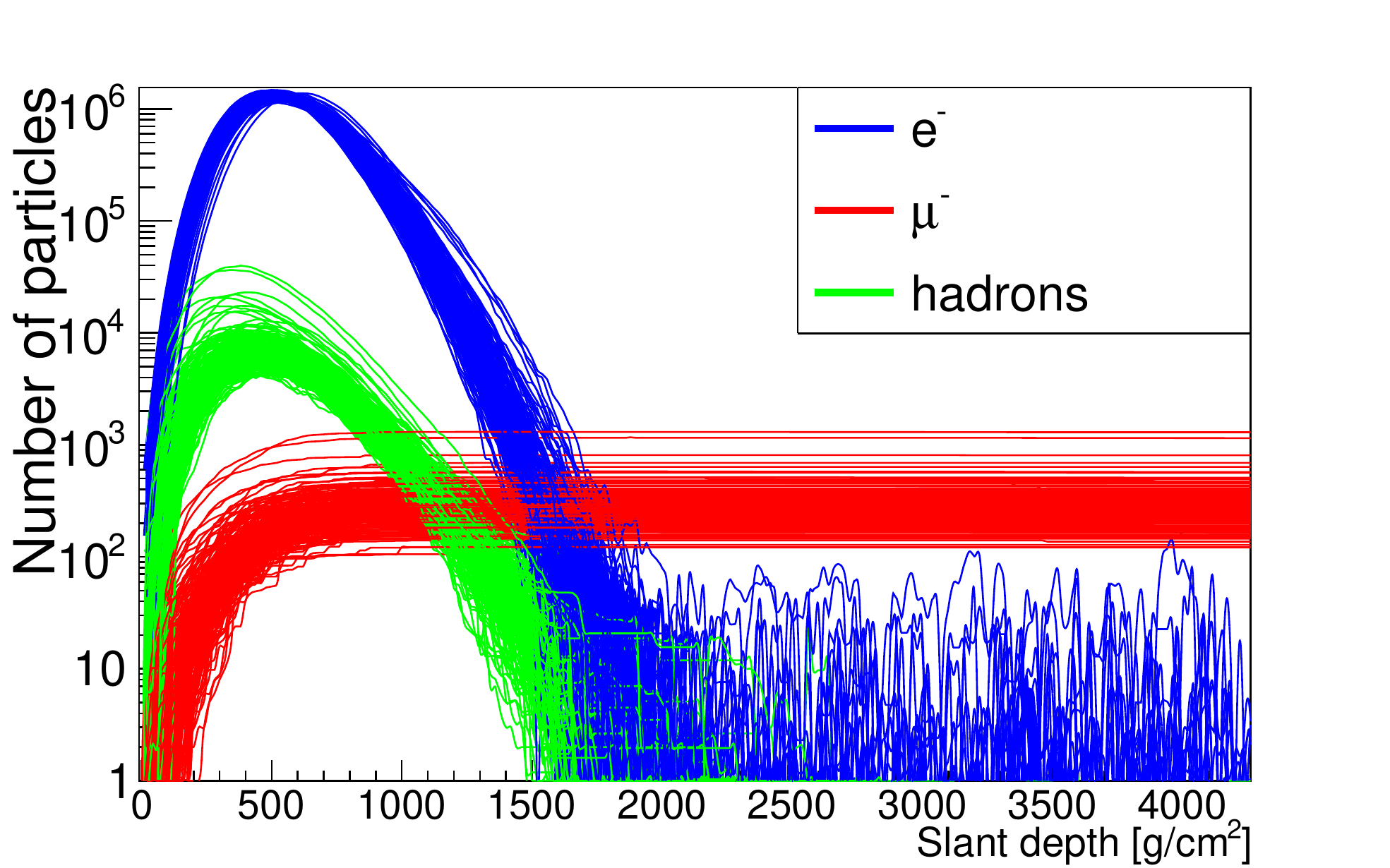}
  \end{subfigure}
  \begin{subfigure}{.5\textwidth}
    \centering
    \includegraphics[width=8cm]{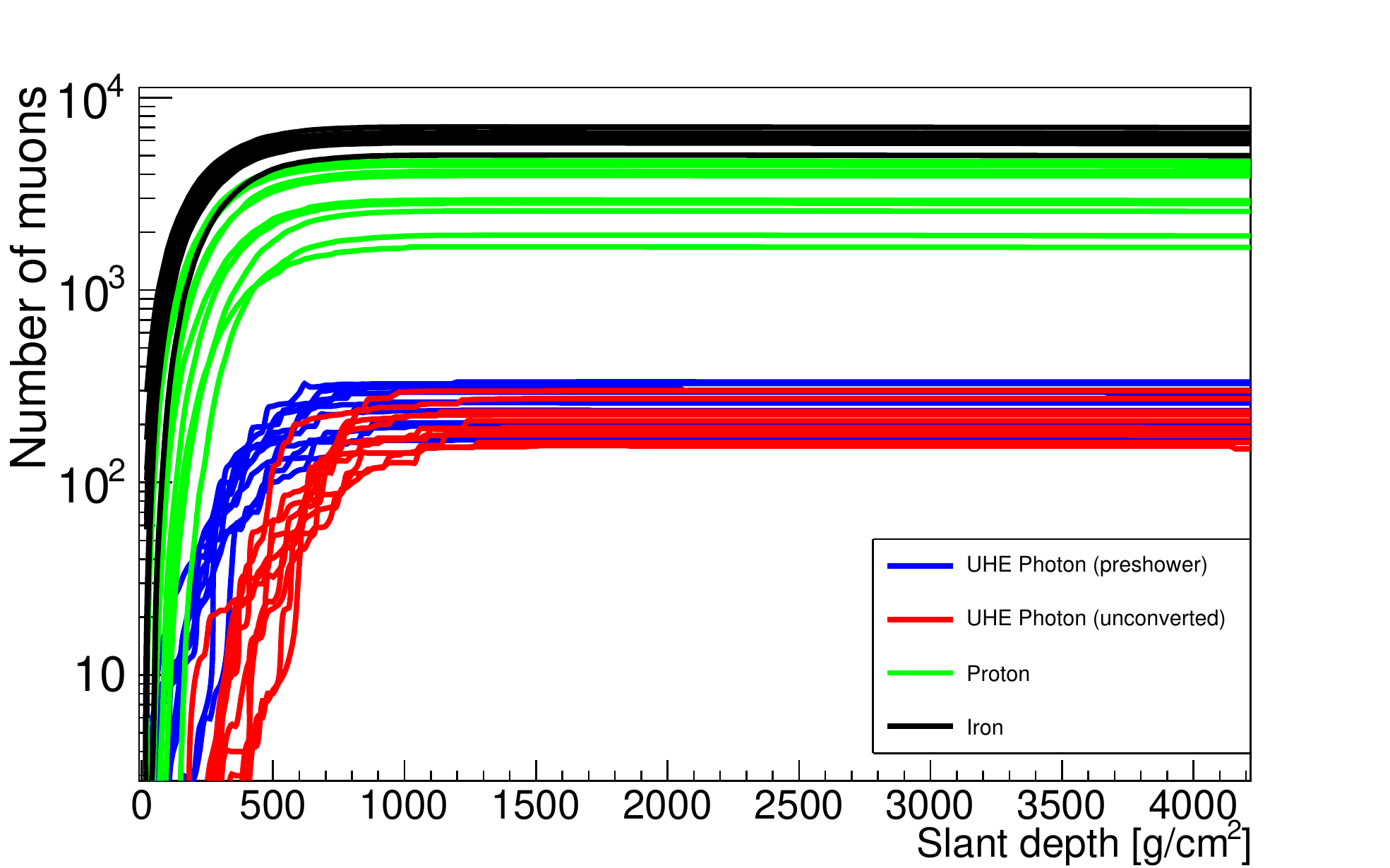}
  \end{subfigure}
  \caption{\textit{Left panel}: Longitudinal profiles of charged particles for 200 air showers generated via preshower effect of a 40 EeV photon coming from the nearly horizontal direction. Altitude is given in slant depth. \textit{Right panel}: Longitudinal profile of the muonic component for different 40 EeV primaries. NOTE: the location of the CTA-North site at 2200 m is equivalent to a slant depth of 4550 $\mathrm{g/cm^{2}}$ at $\theta=80^{\circ}$.}
  \label{fig:muon}
\end{figure*}%

The CR background was simulated for a pure composition of protons, for the maximum impact distance $R_{\mathrm{max}}=1300$ m, a viewing cone angle $\alpha_{\mathrm{CR}} = 5^{\circ}$, and for an energy range extending from 10 TeV to 10 EeV (13 energies simulated as shown in the first row of Table \ref{tab:trigger}). The simulated differential spectrum of CRs takes into account the steepening in the knee region ($\sim 3$ PeV) and the hardening around the ankle ($\sim 3$ EeV), and is described as follow:

\begin{equation}
    \frac{dN_{\mathrm{CR}}(E)}{dE}=N_{0}\left(\frac{E}{E_{0}}\right)^{-\Gamma}, \label{current_rel1}
\end{equation}
expressed in $\mathrm{m^{-2}s^{-1}sr^{-1}TeV^{-1}}$(see footnote\footnote{$\{N_{0},E_{0},\Gamma\}$ is $\{10.9\times10^{-2},1\:\mathrm{TeV},2.75\}$ for $E<3$ PeV \cite{wiebel94}; $\{2.98\times10^{-11},3\:\mathrm{PeV},3\}$ for $3\:\mathrm{PeV} \le E < 3\:\mathrm{EeV}$; and $\{2.98\times10^{-20},3\:\mathrm{EeV},2.75\}$ for $E\ge3$ EeV.} for the set of parameters used at different energy ranges). Figure \ref{fig:trigger} and the third row of Table \ref{tab:trigger} show the differential spectrum of expected CRs for an observation time $\Delta t$ of 30 hours:

\begin{equation}
    \frac{dN_{\mathrm{CR,exp}}}{dE}=A\times \Delta t \times \Omega \times \frac{N_{\mathrm{trig}}(E)}{N_{\mathrm{sim}}(E)}\frac{dN_{\mathrm{CR}}(E)}{dE}, \label{eqn:Nexp}
\end{equation}
where $A=\pi R_{\mathrm{max}}^{2}$ is the simulated area in the plane perpendicular to the shower direction, $\Omega = 2\pi(1-\mathrm{cos}(\alpha_{CR}))$ is the solid angle and $N_{\mathrm{trig}}(E)/N_{\mathrm{sim}}(E)$ is the fraction of CR events of energy $E$ triggering the array. For each energy, we simulated $N_{\mathrm{sim}}(E)=1500$ showers and the number $N_{\mathrm{trig}}(E)$ which triggered the array is given in the second row of Table \ref{tab:trigger}. The number of expected CR events above an energy $E$ is simply obtained by integrating Equation \ref{eqn:Nexp} over the appropriate energy range and considering the differential CR spectra previously described. The results are shown in the last row of Table \ref{tab:trigger}.

In \cite{neronov16}, it was demonstrated that looking at nearly horizontal EASs allows observations of the muonic component, which can be used to discriminate EASs induced by CRs from the ones produced by gamma rays, alongside with the atmospheric depth at which these showers reach their maximum $X_{\mathrm{max}}$. Figure \ref{fig:xmax} (left) shows the $X_{\mathrm{max}}$ distributions for preshowers and unconverted UHE photons. As seen in this plot, EASs initiated by preshowers tend to reach their maximum higher in the atmosphere than the ones produced by unconverted UHE photons. This is largely due to the first interaction point located higher (up to several thousands of kilometers above the atmosphere, as shown in Figure \ref{fig:xmax} (right)) and to the LPM effect which partly suppress the cross-section of bremsstrahlung and pair production. This forces the showers to develop on much larger distances. In the early stages of air shower development, Cherenkov radiation is predominantly emitted by the electromagnetic component. However, due to ionization and bremsstrahlung emission, electrons and positrons suffer significant energy losses and as a consequence, the Cherenkov light emitted by these particles drops drastically once the EAS maximum is reached. On the other hand, muons tend to lose energy through the same processes on much larger distances, resulting in accumulation of these particles as the EAS development progresses and in the existence of a muon plateau.

\begin{figure}[!b]
  \centering
  \includegraphics[width=7.5cm]{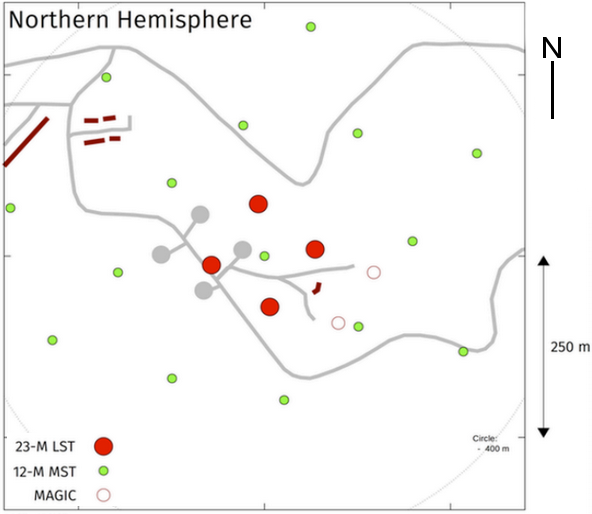}
  \caption{Lay-out of CTA-North, taken from \cite{cta11} and \textit{https://www.cta-observatory.org/about/array-locations/la-palma/}. Since simulations were performed, the planned layout for CTA-North was slightly modified and is different from the one presented in this plot.}
  \label{fig:lapalma}
\end{figure}%

\begin{figure*}[!ht]
  \begin{subfigure}{.5\textwidth}
    \centering
    \includegraphics[width=6cm]{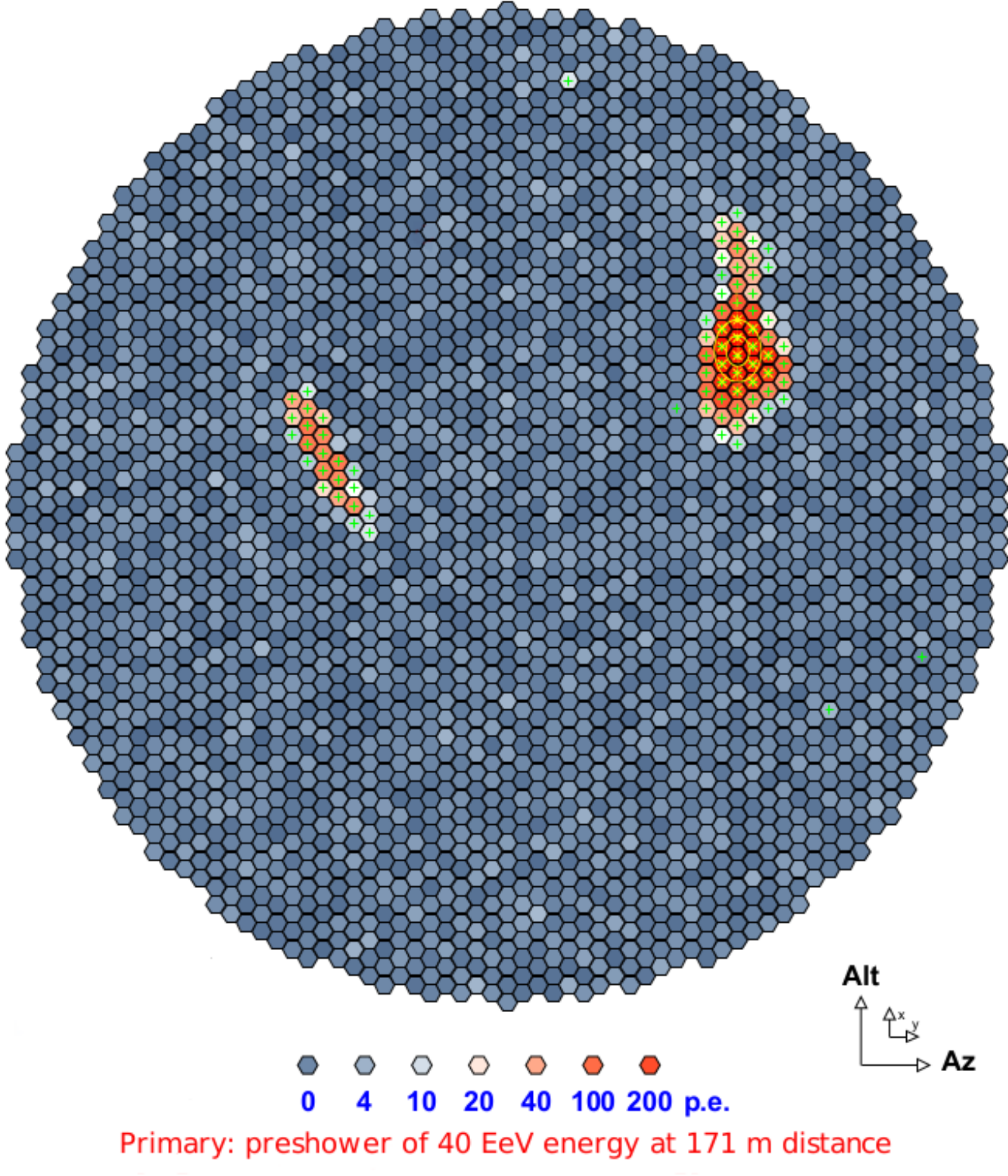}
  \end{subfigure}
  \begin{subfigure}{.5\textwidth}
    \centering
    \includegraphics[width=6cm]{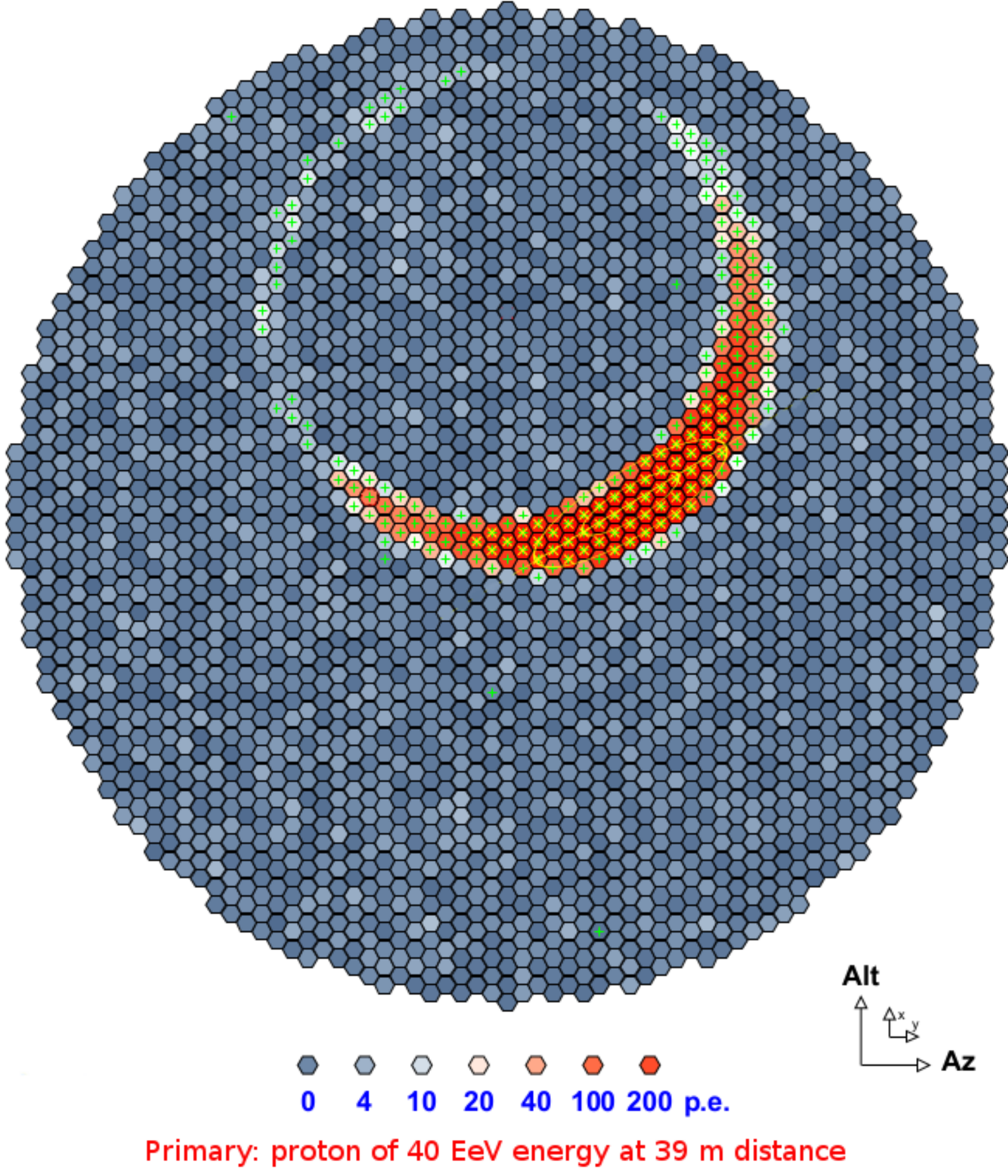}
  \end{subfigure}
  \caption{Example of camera images for preshower effect (left) and proton (right) primaries with $E = 40$ EeV, $\theta=80^{\circ}$ and $\phi=180^{\circ}$. Taken from \cite{almeida17}. }
  \label{fig:images}
\end{figure*}%
\begin{figure*}[!ht]
  \begin{subfigure}{.5\textwidth}
    \centering
    \includegraphics[width=7.5cm]{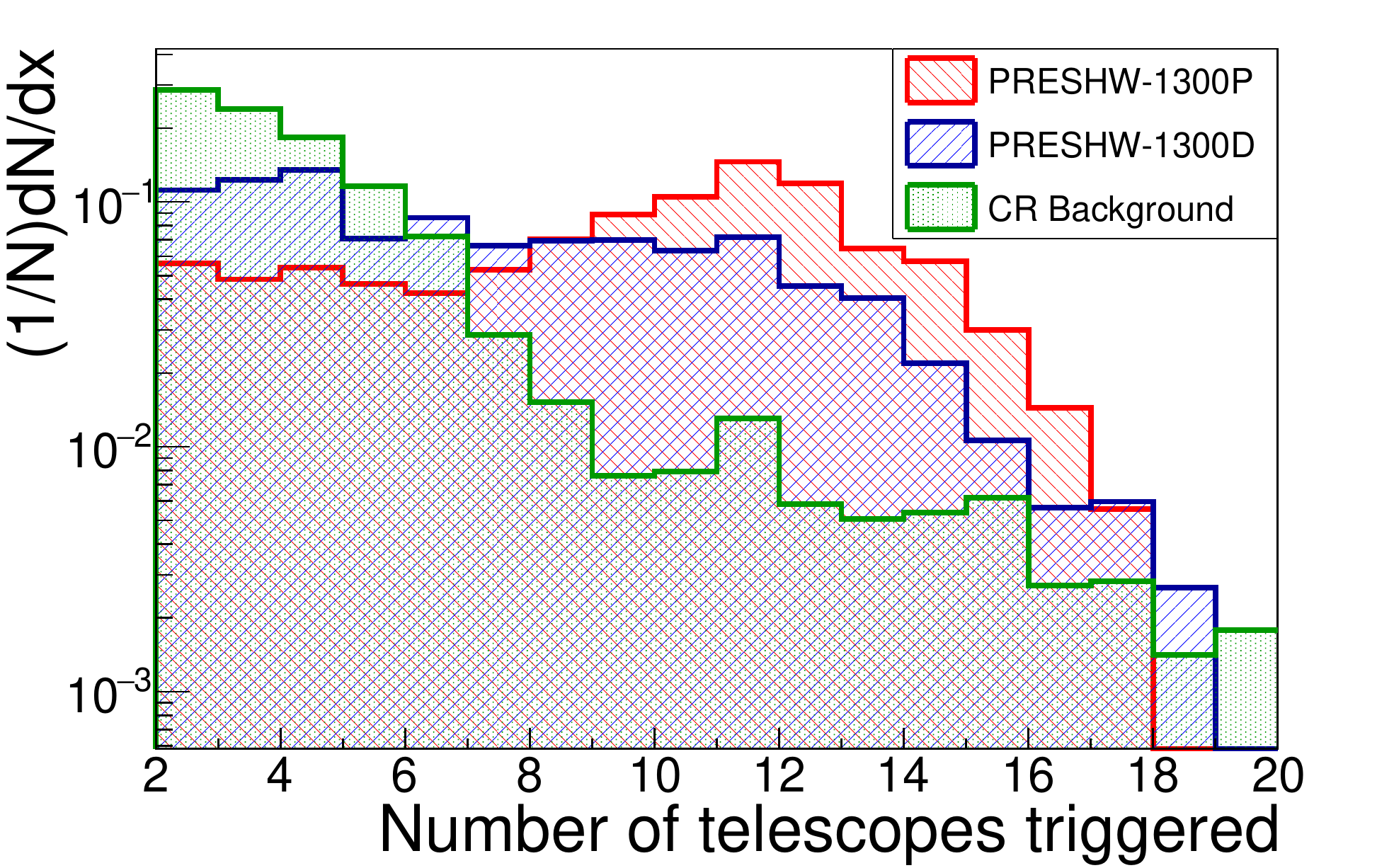}
  \end{subfigure}
  \begin{subfigure}{.5\textwidth}
    \centering
    \includegraphics[height=4.5cm]{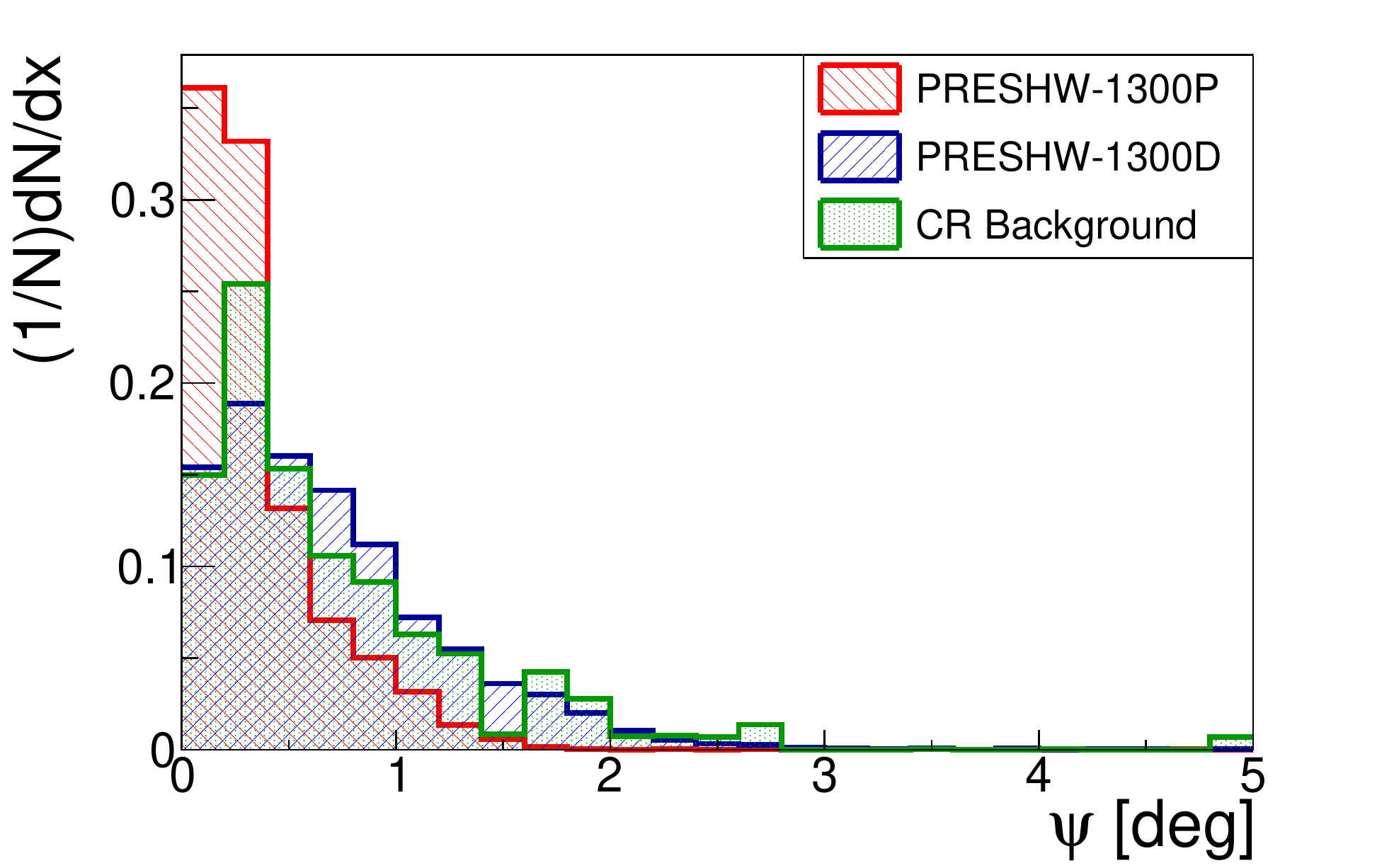}
  \end{subfigure}
  \caption{\textit{Left panel}: Normalized distributions of the number of telescopes triggered by air showers generated by the CR background (with $R_{\mathrm{max}} = 1300$ m) and by the preshower effect of different parameters. \textit{Right panel}: Normalized distributions of the difference between reconstructed and simulated directions for the same air showers.}
  \label{fig:dist}
\end{figure*}%
Such EAS development behavior was shown in \cite{neronov16} to be the characteristic of gamma rays, iron and proton-initiated showers and Figure \ref{fig:muon} shows that it is also the case for showers resulting from the preshower effect. Although muons are much less efficient at emitting Cherenkov radiation than electrons, their preponderance in the late stage of the EAS development as well as the fact that muons emits such radiation closer to the surface make the muonic component detectable by telescopes looking over the horizon.

In the next sections, the sets of parameters used for different preshower scenarios are specified in the \textit{PRESHW-1300Y} format, where \textit{Y} refers to the nature of the source (\textit{P} if it is a point source with $\alpha_{\mathrm{preshw}} = 0^{\circ}$ or \textit{D} if it is a diffuse source with $\alpha_{\mathrm{preshw}} = 5^{\circ}$). In these scenarios, the zenith and azimuth angles are set to $\theta=80^{\circ}$ and $\phi=180^{\circ}$, respectively, with energy $E = 40$ EeV.

\subsection{CTA's detectors response}

The detection of the Cherenkov light is performed by the cameras of IACTs. CORSIKA's output from preshower effect and CR background EASs is then piped into \textit{sim\_telarray} \cite{bernlohr08}, a software that simulates the detectors response of the telescopes array, including the cameras electronics, the optical-ray tracing and the recording of photons in the photomultiplier tubes. The \textit{production-I} settings allow us to define the properties of each type of the telescopes planned to be part of the array in La Palma (4 large-sized telescopes with a field of view (FOV) of $4.3^{\circ}$ and 15 medium-sized telescopes with a FOV of $8^{\circ}$). The area covered by the array is approximately 0.6 $\mathrm{km^{2}}$, which must be compared to the simulated area that is obtained by projecting the circle of radius $R_{max}$ on the ground and perpendicularly to the circle. The ellipse formed on the ground has an area of 30.6 $\mathrm{km^{2}}$ for $R_{max}=1300$ m. The number of pixels needed for each telescope to be triggered is set to 3 and the number of telescopes needed to trigger the system is left to the default value of 2. A background of photo-electrons (p.e.) is simulated at 122 MHz (0.122 p.e/ns) for each telescope in order to account for the night-sky background (NSB). The telescopes pointing direction is set to the $\theta=80^{\circ}$, $\phi=180^{\circ}$ direction with an offset of $0.5^{\circ}$ in zenith angle for the preshower simulations. Such offset is introduced to simulate a wobble mode of observation, as documented and used in the case of the MAGIC telescope. Such observation mode consists in pointing the telescopes slightly off the source. By doing so, both background and signal can be measured with the same systematics as the background is therefore measured with the same zenith angle and the same weather conditions \cite{bretz05}.

Two examples of images formed on the cameras by Cherenkov radiation are shown in Figure \ref{fig:images} for proton and preshower primaries (see also \cite{almeida17}). One very unique characteristic of nearly horizontal air showers is the presence of muon rings (right image) due to the dominating muonic component of the air showers as previously mentioned. Another interesting feature is the presence of multiple images in single cameras in the case of preshower primaries (left image). The V-shape formed by the two images on the IACTs cameras by EAS initiated by multiple photons can be used as an additional discrimination criteria thanks to the angular resolution of CTA cameras \cite{sitarek19}. Such a property would constitute a powerful tool to identify preshowers in an efficient way. These images can be described with a set of parameters that can be used to discriminate images from CR background and preshowers. Further details are given in section \ref{sec-variables}.

Figure \ref{fig:dist} (left) shows the number of telescopes triggered by the simulated CR background as well as by point and diffuse sources of preshower. As can be seen, the CR background tends to trigger fewer telescopes than preshowers. This difference can be explained by the fact that the CR background observed by the array is dominated by air showers generated by low-energy particles ($\sim$TeV) which contain muons of lower energy than in the case of preshower-induced air showers. Hence, the amount of Cherenkov light produced by these muons is not sufficient to irradiate and trigger a large number of telescopes. It should be noted that the few CR events that trigger a large number of telescopes (15 and above) correspond to the upper end of our simulated energy range ($10$ EeV) described in the previous section. The right panel of Figure \ref{fig:dist} shows the discrepancy between the simulated $\vec{V}_{\mathrm{sim}}$ and reconstructed $\vec{V}_{\mathrm{rec}}$ directions via the $\psi$ angle\footnote{\begin{equation} \psi = \arccos{(\vec{V}_{\mathrm{sim}}.\vec{V}_{\mathrm{rec}})} \end{equation}}. We note that the reconstruction yields better results in the case of point sources ($\psi_{\mathrm{median}}^{\mathrm{PRESHW-1300P}} = 0.28^{\circ}$) as the difference between the air shower  axis and the pointing directions of the telescopes is less significant than in the case of diffuse sources ($\psi_{\mathrm{median}}^{\mathrm{PRESHW-1300D}} = 0.59^{\circ}$ and $\psi_{\mathrm{median}}^{\mathrm{CR}} = 0.72^{\circ}$). As the angular resolution is smaller than the FOV of the telescopes, no events are lost via the reconstruction procedure.

\section{Preshower/CR background separation}

In observational astronomy and for any channel of observation, the sensitivity of a telescope is directly related to the efficiency with which one can extract signal from background. In the case of gamma-ray astronomy, the background is composed of EAS generated by an isotropic flux of CRs. In order to correctly discriminate these EAS against the ones produced by preshowers, a multivariate analysis based on the characteristics of the camera images has to be performed. Although several methods exist (as discussed in \cite{bock04}), the use of \textit{Boosted Decision Trees} (BDT) has proven to be the most efficient at improving the background rejection in leading gamma-ray experiments such as H.E.S.S. \cite{ohm09} and MAGIC \cite{albert08}. In this section, we briefly review the basic elements of the BDT analysis developed in the TMVA package \cite{hocker07} and we discuss the variables used to obtain the best preshower/CR background separation for different preshower scenarios. 

\subsection{Building of a decision tree}

\begin{figure}[!t]
  \centering
  \includegraphics[width=7.5cm]{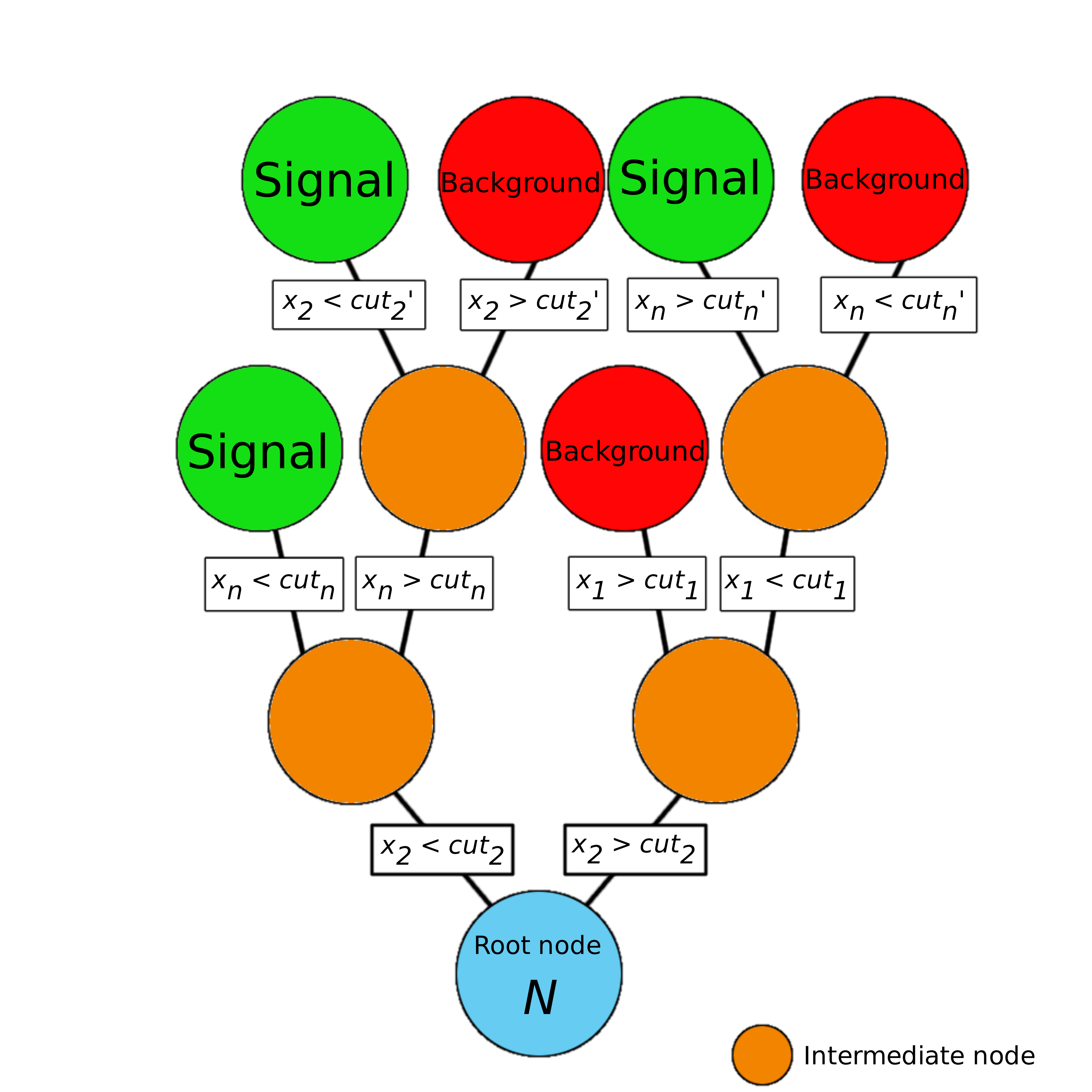}
  \caption{Schematic view of a decision tree.}
  \label{fig:tree}
\end{figure}%

\begin{figure*}[ht]
  \centering
  \includegraphics[width=13.7cm]{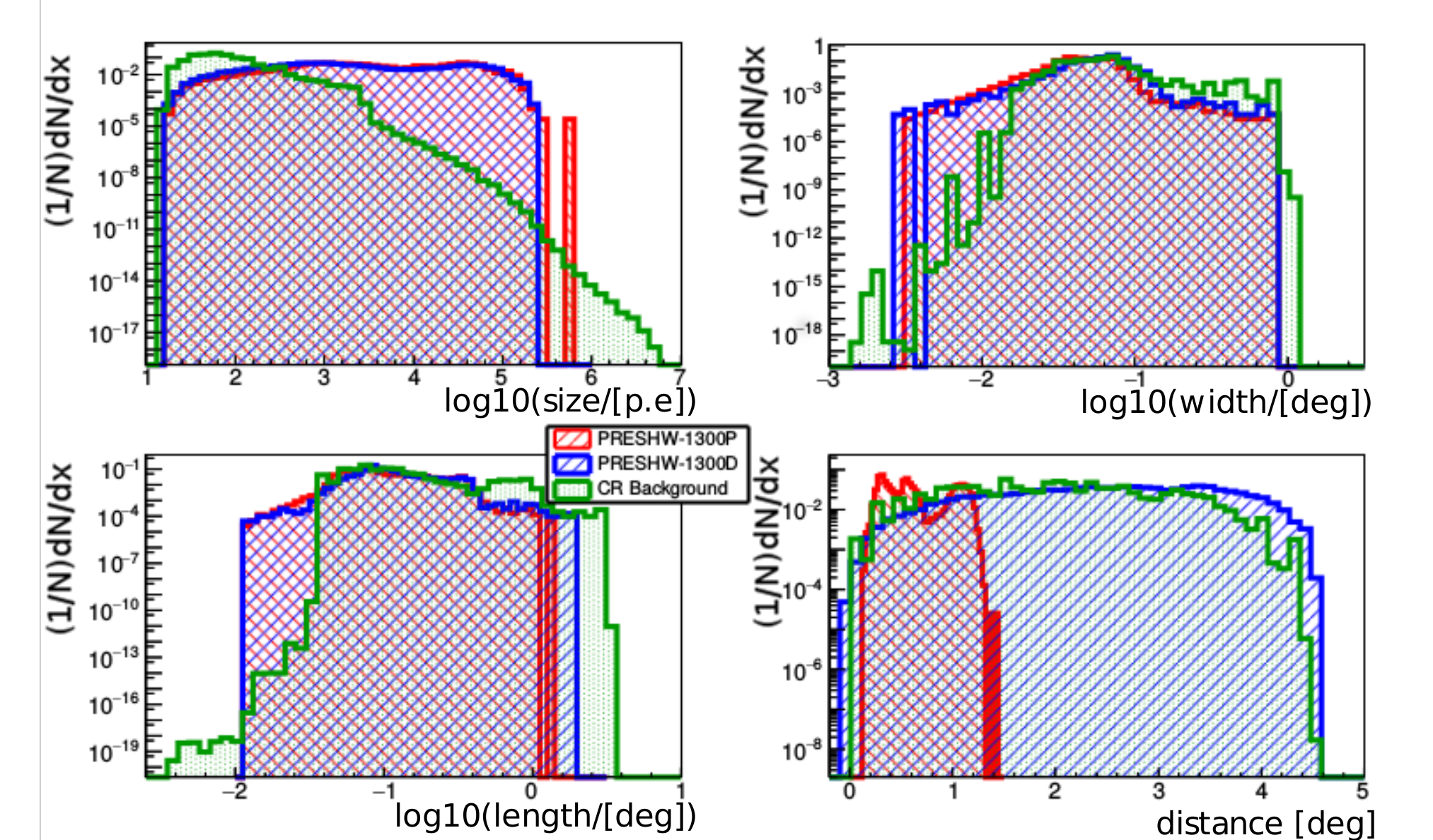}
  \caption{Normalized distributions of the four Hillas parameters used in the multivariate analysis for both preshower and CR background simulations.}
  \label{fig:training_variables}
\end{figure*}%

\begin{figure*}[!ht]
  \begin{subfigure}{.5\textwidth}
    \centering
    \includegraphics[width=7.5cm]{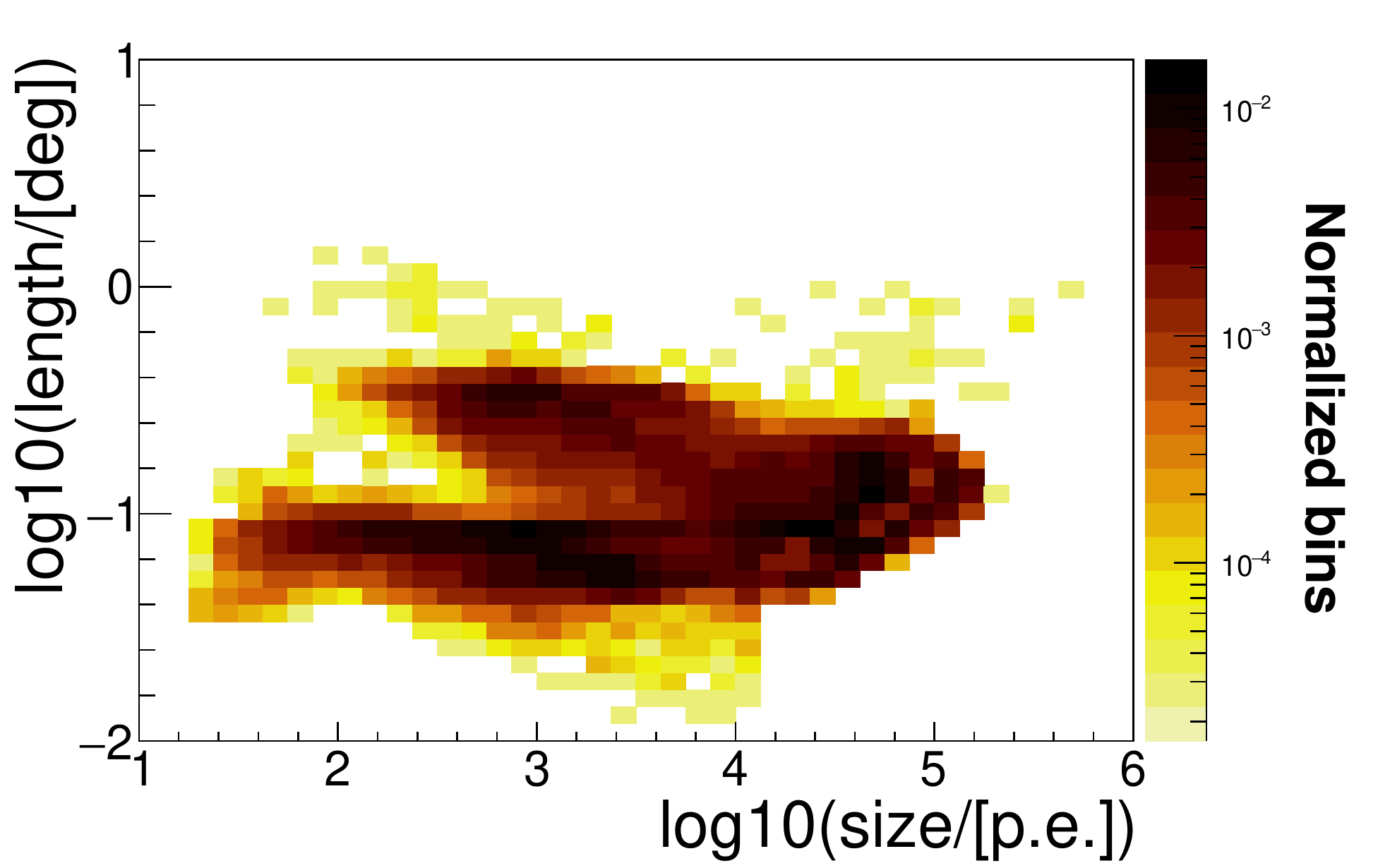}
  \end{subfigure}
  \begin{subfigure}{.5\textwidth}
    \centering
    \includegraphics[width=7.4cm]{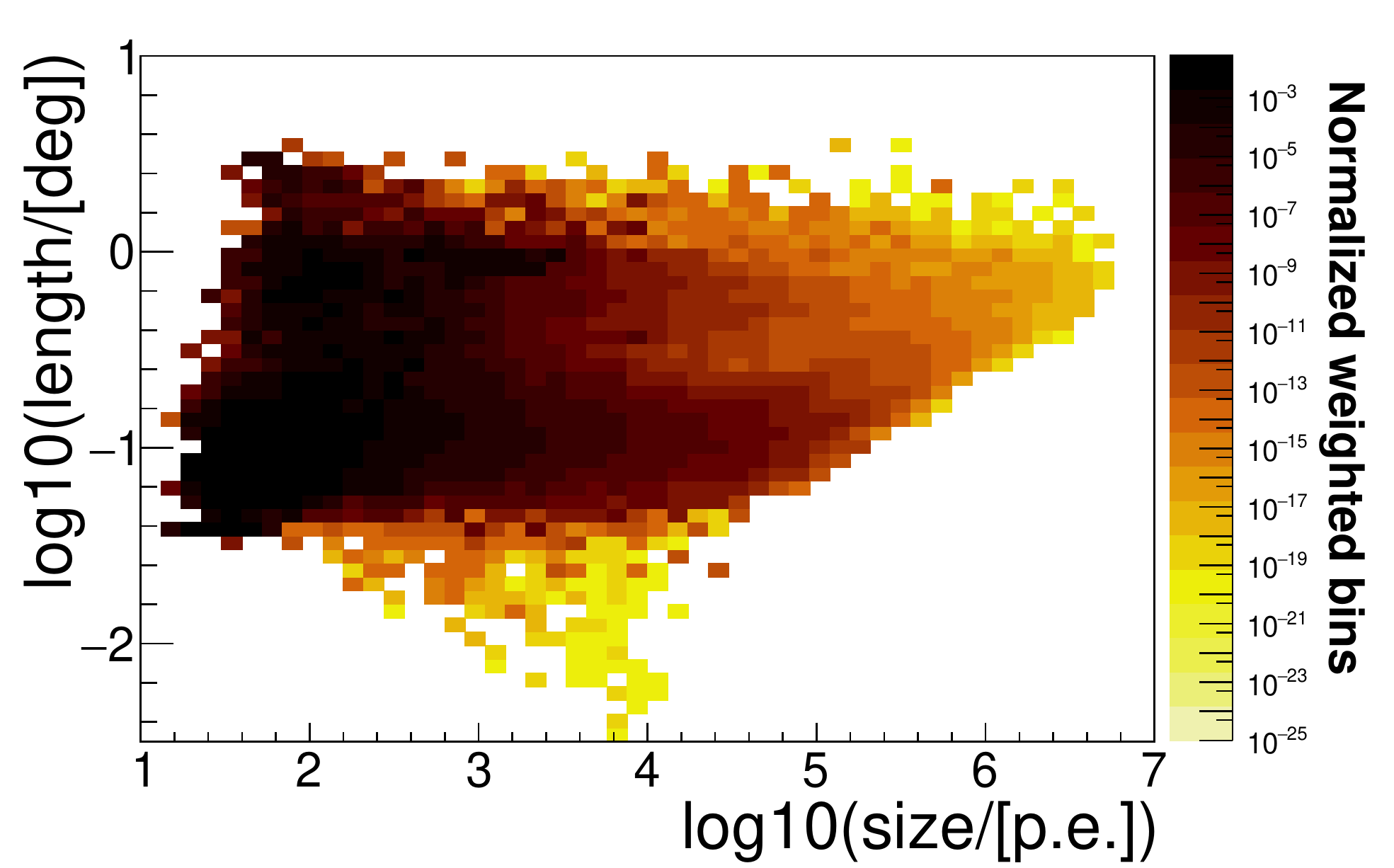}
  \end{subfigure}
  \caption{Scatter plots of \textit{size} and \textit{length} parameters for PRESHW-1300P (left) and for the CR background (right). The CR background plot is weighted by the CR spectrum index.}
  \label{fig:scatter}
\end{figure*}%

The basic principle of a decision tree is to classify events by applying successive cuts on the variables that characterize them. In order to grow such a tree, one needs to define a training sample composed of events of known nature (signal or background). Each one of these events is characterized by a set of \textit{n} variables $\{x_{1},x_{2},..,x_{n}\}$. The growing of the tree starts at the \textit{root node}, where all the events of the training sample are contained. A first binary split is applied to the dataset based on the variable and its corresponding cut value that produce the best signal/background separation. The same operation is then repeated for each one of the two resulting subsets until one of the criteria set to stop the building of the tree is met (maximum depth of the tree, minimum number of events in a node, etc...). Each one of the last nodes, also called \textit{leaf nodes}, is then labeled \textit{signal} or \textit{background} depending on the class of the majority of events contained in it (see Figure \ref{fig:tree}). Individual trees are very sensitive to fluctuations in the training sample and therefore constitute weak classifiers. To overcome this problem, multiple trees are trained into a \textit{forest} and misclassified events gain a \textit{boost weight} that increase their importance when the next tree is grown: this is the \textit{boosting} procedure and it is performed by the \textit{AdaBoost} method found in the TMVA package.

\subsection{Hyperparameters}

One of the advantages of the BDT method lies in the fact that very little tuning of the method's parameters (also called \textit{hyperparameters}) is required. In this section, we briefly review the hyperparameters that characterize the BDT method and the values that were used.

The number of trees (\textit{NT}) to be grown can be determined by looking at the fraction of misclassified events for each newly constructed tree. As more trees are grown, this fraction should converge towards 0.5, the threshold value for which no more additional separation power can be obtained from the data. $NT=800$ provides both a good separation power and a value above which trees do not significantly improve it. Another parameters that has an impact on the separation power is the maximum depth (\textit{MaxD}) of the trees. A good compromise is found with $MaxD=3$, preventing overtraining (an issue that arises when a classifier becomes too sensitive to statistical fluctuations and therefore, does not perform well on independent samples) but also providing an efficient signal/background separation. The minimum fraction of training events contained in a node (\textit{MinNodeSize}) must also be carefully set. A high value would leave leaf nodes with a too high percentage of misclassified events while a low value would lead to overtraining. An acceptable balance is achieved by using the value of $2.5\%$. The number of cuts \textit{nCuts} used to find the optimal node splitting for each variable was set to 500 in the case of point sources of UHE photons and to 300 in the case of diffuse sources. Finally, the \textit{Gini coefficient} \cite{gini1912} was selected as a separation criterion to find the optimal cut value at each node by comparing the background and signal distributions of the training variables.

\subsection{Classification variables} \label{sec-variables}

Cherenkov light emitted by secondary particles in the EASs is recorded by the cameras of IACTs and forms images which can be characterized by a set of geometrical parameters called Hillas parameters \cite{hillas85}. These parameters strongly depend on the nature and the energy of the primary particle that generated the EAS and can therefore be used to discriminate signal and background events. The parameters selected for our analysis are chosen based on their importance (also called ranking) when running the BDT analysis: the more a given variable is used to build decision trees, the higher its ranking. Figure \ref{fig:training_variables} shows the distributions of four Hillas parameters deemed the most important to the classification process of CR background and preshowers created in different scenarios. 

The \textit{size} parameters corresponds to the number of photo-electrons (p.e.) created by the Cherenkov light in the cameras pixels. It is strongly correlated to the energy of the primary which initiated the EAS and on the distance between the detectors and the maximum of that EAS, i.e. $X_{\mathrm{max}}$. As Figure \ref{fig:preshw} (right) shows, the preshower effect on a UHE photon produces multiple photons in the EeV regime which initiate EASs. These preshower-initiated EASs reach their maximum development at a much closer distance to the IACTs than the ones produced by the CR background. However, the most energetic CRs contain far more muons (see right panel of Figure \ref{fig:muon})  and may generate far brighter images, as shown on the top-left plot of Figure \ref{fig:training_variables}. 

One common feature is the existence of ring-shaped images which are characteristic of the muonic component. However, images formed by CR-induced EAS tend to be more irregular and larger compared to the one of EAS produced by photons. Consequently, the \textit{length} and \textit{width} distributions reach larger values in the case of proton-induced showers.

The \textit{distance} parameter characterizes the angular distance between the center of the camera and the simulated source position. The bottom-right panel of Figure \ref{fig:training_variables} shows a clear difference in the distribution of the distance parameter between point sources (\textit{PRESHW-1300P}) and diffuse sources (\textit{PRESHW-1300D} and CR background), with diffuse sources -- and especially of preshowers -- showing a more extended distribution of this parameter. This is due to the fact that by nature, diffuse sources have a larger spread of angle distribution between the axis of the EAS and the one of the telescopes.

\begin{figure*}[!t]
  \begin{subfigure}{.5\textwidth}
    \centering
    \includegraphics[width=7.5cm]{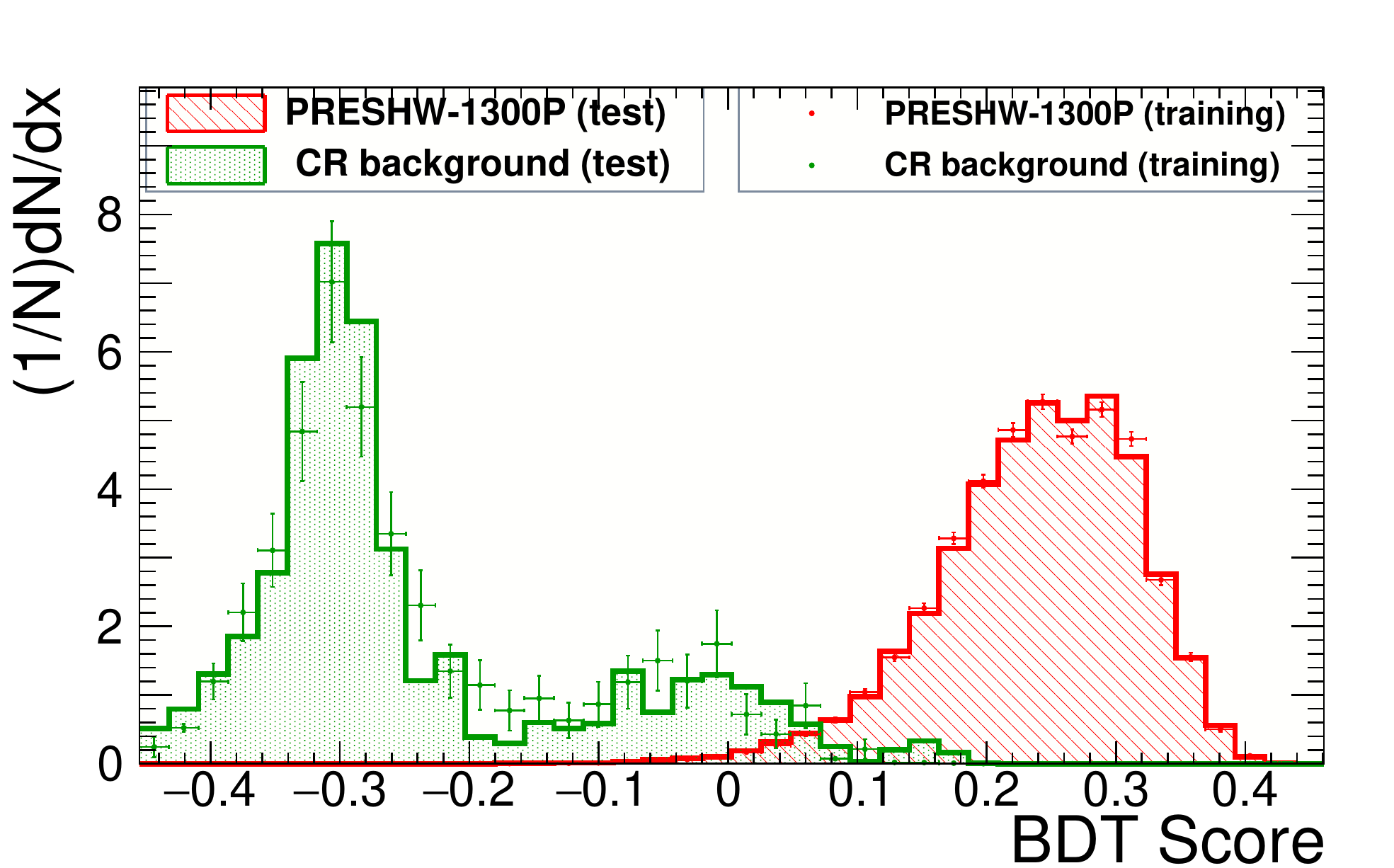}
  \end{subfigure}
  \begin{subfigure}{.5\textwidth}
    \centering
    \includegraphics[width=7.5cm]{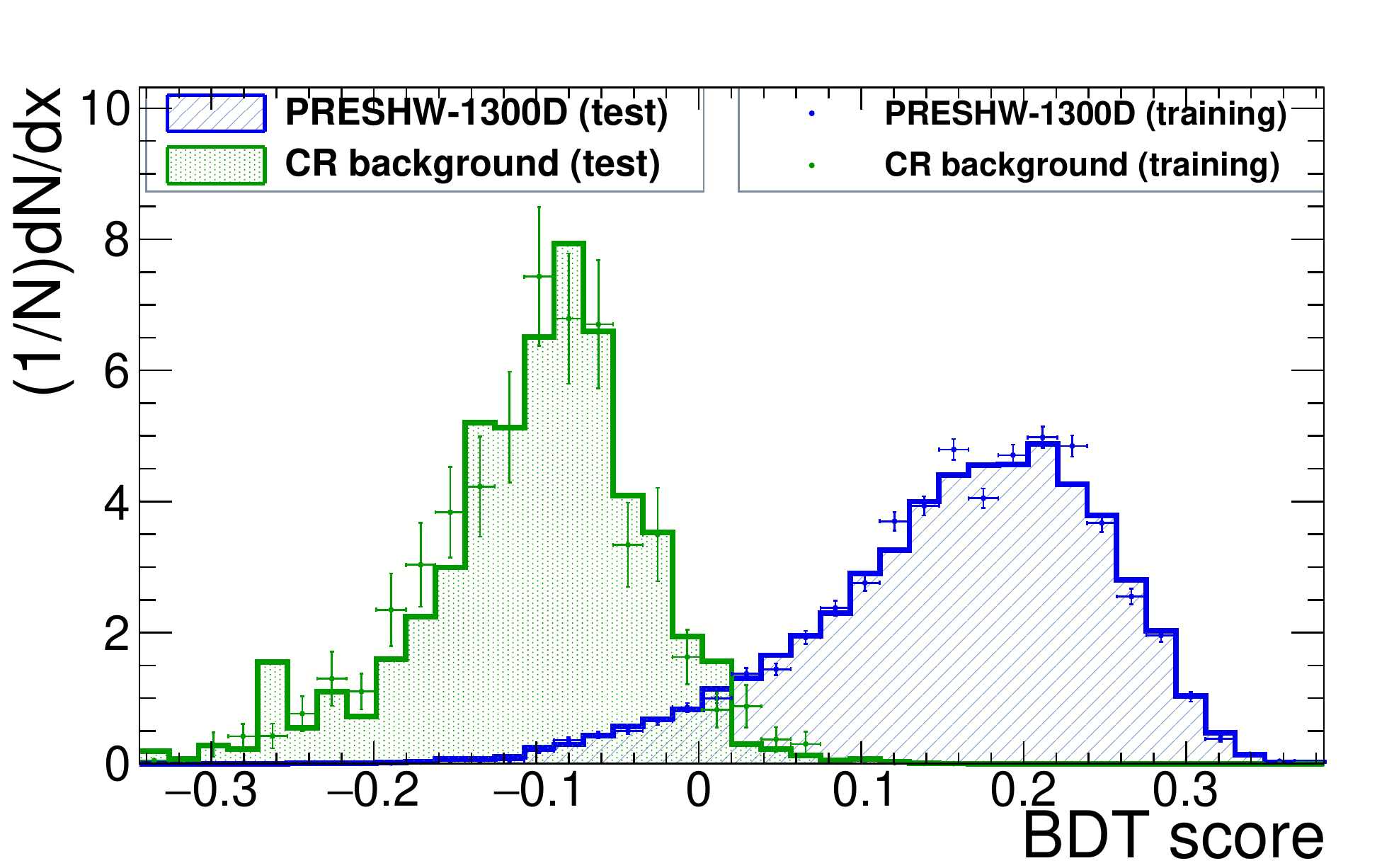}
  \end{subfigure}
  \caption{BDT score distributions of the testing and training samples of CR background and preshower events.}
  \label{fig:BDTdist}
\end{figure*}%

\begin{figure*}[!t]
  \begin{subfigure}{.5\textwidth}
    \centering
    \includegraphics[width=8cm]{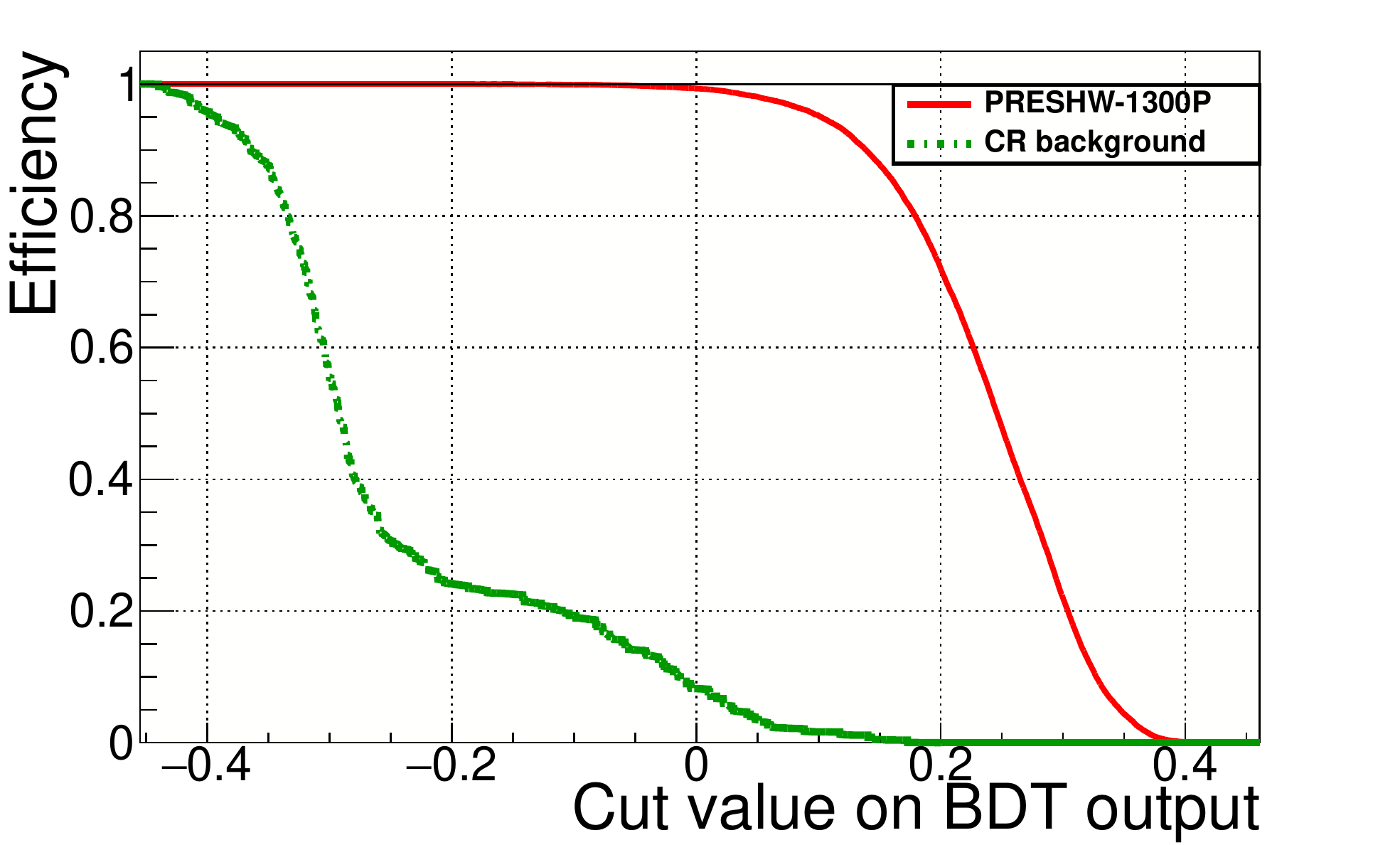}
  \end{subfigure}
  \begin{subfigure}{.5\textwidth}
    \centering
    \includegraphics[width=8cm]{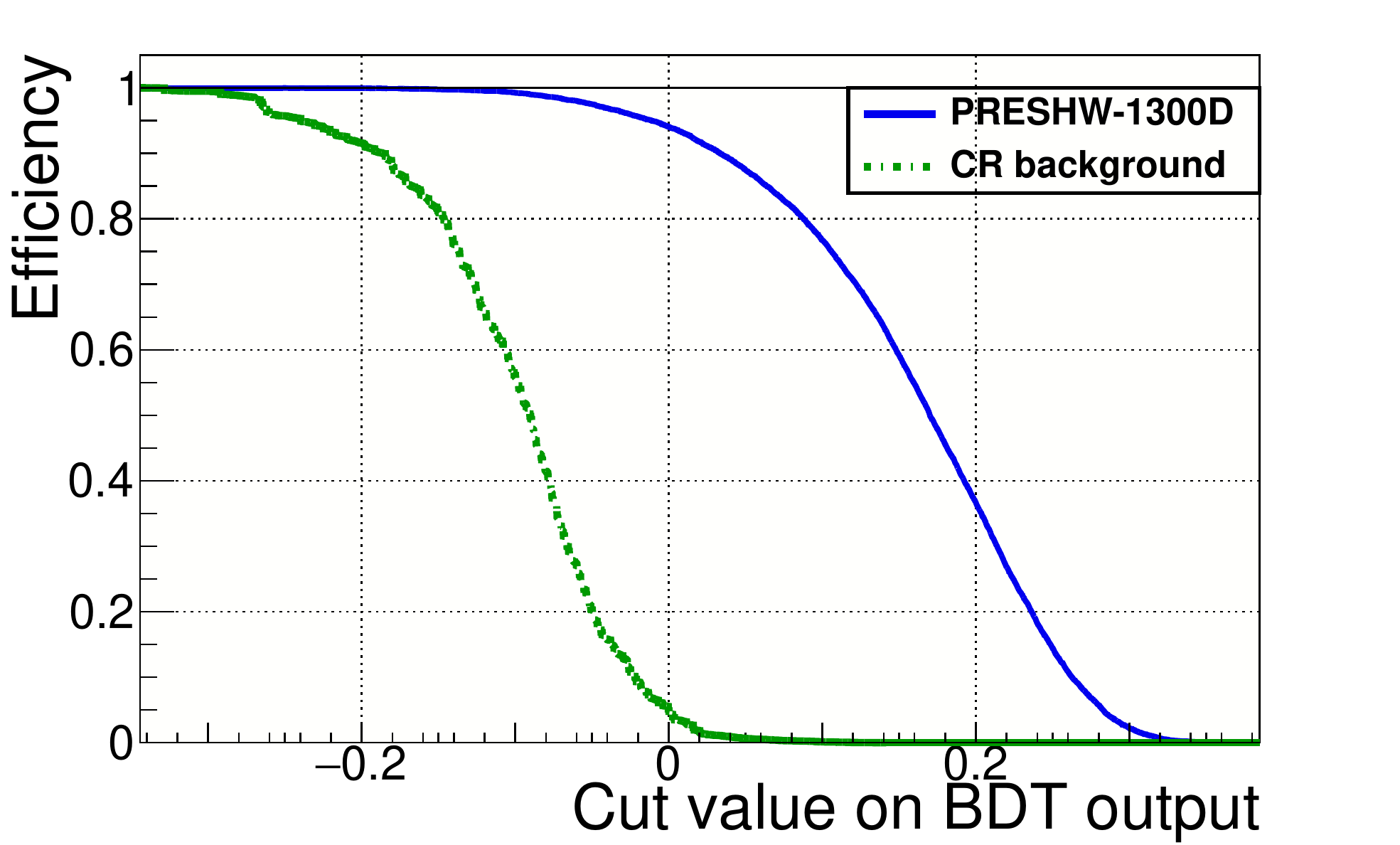}
  \end{subfigure}
  \caption{Efficiencies obtained for CR background and preshower events using test samples as a function of cut value applied to distributions of Figure \ref{fig:BDTdist}.}
  \label{fig:BDTefficiencies}
\end{figure*}%

\begin{table*}[ht]
\centering
\begin{tabular}{clllll}

& Best MCC & BDT cut value & Sig. Eff. $\epsilon_{\mathrm{BDT}}$ & Back. Con. & Sig. Pur.\\ \hline \hline
\multicolumn{1}{c}{PRESHW-1300P}&\multicolumn{1}{c}{0.96}&\multicolumn{1}{c}{0.08}&\multicolumn{1}{c}{96.6\%}&\multicolumn{1}{c}{2.1\%}&\multicolumn{1}{c}{97.9\%}\\  
\multicolumn{1}{c}{PRESHW-1300D}&\multicolumn{1}{c}{0.89}&\multicolumn{1}{c}{0.07}&\multicolumn{1}{c}{83.8\%}&\multicolumn{1}{c}{0.4\%}&\multicolumn{1}{c}{99.6\%}\\
\hline \hline
\end{tabular}
\caption{Summmary of the BDT classification results obtained for different parameter sets of preshower simulations. The signal efficiency, the background contamination as well as the signal purity for the best MCC value obtained are shown.}
\label{tab:eff}
\end{table*}

Figure \ref{fig:scatter} shows the scatter plots of \textit{size} vs. \textit{length} for the \textit{PRESHW-1300P} scenario and the CR background. In the former case, a wide variety of images are observed, from the small but bright ones ($\log_{10}(size) > 4$) to the typical muon rings ($\log_{10}(size) < 4$ and $\log_{10}(length) > -0.8$), larger in sizes but also dimmer. The scatter plots obtained for the CR background are consistent with observations from MAGIC \cite{magic18} in a 30 hours time period in the darkest region, where events of energies in the $\sim$ TeV-PeV range are observed. The muon rings of different intensities are also noticeable but the vast majority of the images are dim and small.

\subsection{Results from multivariate analysis}

\begin{figure}[!b]
  \centering
  \includegraphics[width=8cm]{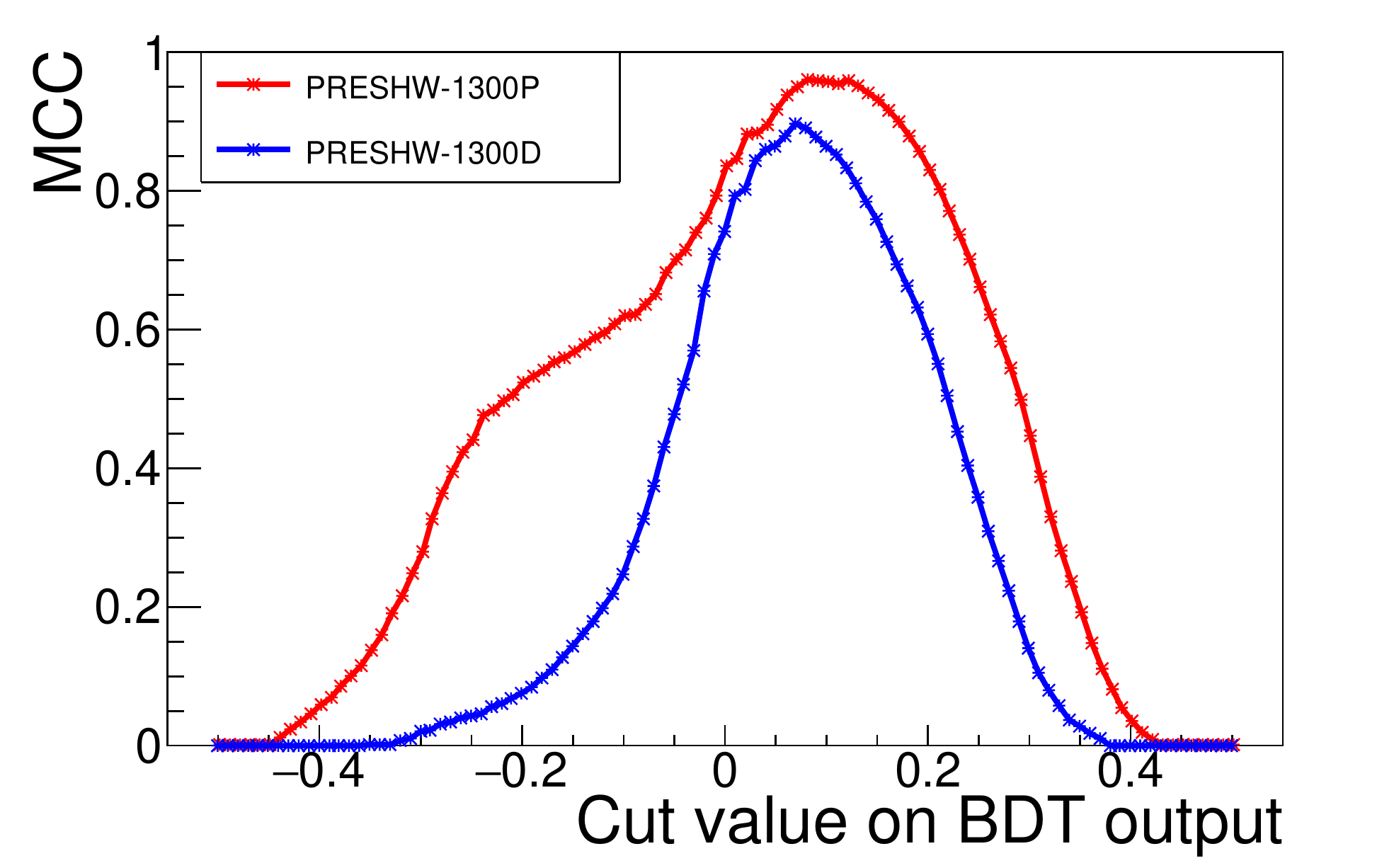}
  \caption{MCC values obtained as a function of the cut applied on the BDT distributions of training samples for different preshower scenarios.}
  \label{fig:mcc}
\end{figure}

The ultimate goal of the BDT method is to assign a score to the training samples such as the separation between signal and background score distributions is maximum. Such BDT score distributions are shown in Figure \ref{fig:BDTdist}. By applying successive cuts on these distributions, one can calculate the evolution of the signal and background efficiencies (as shown in Figure \ref{fig:BDTefficiencies}) and obtain the best \textit{Matthews correlation coefficient} \cite{matthews75} in order to evaluate the quality of the classification of the training sample. It varies from -1 (total contradiction between prediction and observation) to 1 (perfect prediction) and is given by the formula:

\begingroup
\footnotesize
\begin{equation}
     MCC = \frac{TP \times TN - FP \times FN}{\sqrt{(TP+FP)(FP+FN)(TN+FP)(TN+FN)}}, \label{eqn:mcc}
\end{equation}
\endgroup
where \textit{TP} is the number of signal events classified as signal (true positive), \textit{TN} the number of background events classified as background (true negative), \textit{FP} the number of background events classified as signal (false positive), and \textit{FN} the number of signal events classified as background (false negative). Figure \ref{fig:mcc} shows the evolution of this parameter for different cut values performed on the training sample BDT distributions.

In order to evaluate the quality of the trained classifier, it is then used to classify an independent test sample of preshower and CR background events. The optimal cut value previously obtained from the training sample is then applied to the BDT score distribution of the test sample and the final efficiencies are calculated. The same procedure is then applied to all simulation parameter sets and a summary of the results is given in Table \ref{tab:eff}. These results show that the preshowers can be efficiently discriminated from the dominating CR background. In fact, very low contamination from the latter is observed and one could easily obtain completely background free observations with very little loss of preshower efficiency. Another noticeable feature is the fact that diffuse sources of preshowers are not as well discriminated from CR background as preshowers initiated by a point source of UHE photons. Such a discrepancy can be explained by the very similar distributions of the \textit{distance} parameters between the diffuse CR background and the diffuse source of UHE photons. 

\begin{table*}[!ht]
\centering
\begin{tabular}{cll}
 & \multicolumn{1}{c}{PRESHW-1300P} & \multicolumn{1}{c}{PRESHW-1300D} \\ \hline \hline
Aperture [$\mathrm{km^{2}}$] & \multicolumn{1}{c}{3.42} & \multicolumn{1}{c}{0.05} \\ \hline       
SHDM & \multicolumn{1}{c}{$2.7\times10^{-3}$} & \multicolumn{1}{c}{$3.1\times10^{-6}$} \\
GZK & \multicolumn{1}{c}{$4.0\times10^{-5}$} & \multicolumn{1}{c}{$4.7\times10^{-8}$} \\
AUGER$_{\mathrm{diff.}}$ & \multicolumn{1}{c}{$3.3\times10^{-5}$} & \multicolumn{1}{c}{$3.9\times10^{-8}$} \\
\hline \hline
\end{tabular}
\caption{\textit{Diffuse source of UHE photons} -- Summary of the results obtained for a given aperture and associated number of expected preshowers for 30 hours observation time. Different preshower simulation sets are shown. \textit{SHDM} and \textit{GZK} rows correspond to the UHE photons production models referred to in the text while the \textit{AUGER$_{\mathrm{diff.}}$} correspond to the limits put by the Pierre Auger Observatory on the integral diffuse photon flux \cite{rautenberg19}.}
\label{tab:N_E}
\end{table*}

\section{Effective area and event rate calculations}

The effective area corresponds to the area within which the shower core must lie in order for the telescopes to detect the shower. It is a function of the energy E of the primary and of the zenith and azimuth angles, $\theta$ and $\phi$, respectively, and can be expressed as:
\begin{equation}
    A_{eff}(E,\theta,\phi,\epsilon_{\mathrm{BDT}})=\pi R_{\mathrm{max}}^{2}\Omega\frac{N_{\mathrm{trig}}(E,\theta,\phi,\epsilon_{\mathrm{BDT}})}{N_{\mathrm{conv}}}, \label{eqn:effective_area}
\end{equation}
where $R_{\mathrm{max}}$ is the maximum impact distance defined previously, $N_{\mathrm{trig}}(E,\theta,\phi,\epsilon_{\mathrm{BDT}})$ is the number of events that have triggered the telescope array and that have been correctly identified as preshowers -- that is the signal efficiency $\epsilon_{\mathrm{BDT}}$ must be taken into account when calculating this term -- $N_{\mathrm{conv}}$ is the number of UHE photons that converted (taking into account the conversion factor $\epsilon_{\mathrm{conv}}$ of UHE photons at the given direction of observation, it is simply $N_{\mathrm{sim}}\times\epsilon_{\mathrm{BDT}}$ with $N_{\mathrm{sim}}=10000$) and $\Omega$ is the solid angle in steradian ($\Omega=2\pi(1-cos(\alpha))$ for diffuse sources, where $\alpha$ is the viewcone angle previously defined and $\Omega=1$ for point sources as there are no dependence on the solid angle in this case). The results obtained for the effective area in the case of preshowers are shown in the second row of Table \ref{tab:N_E}.    

Because of limitations on the simulated energies of electromagnetic particles within the EASs, these values are conservative. Lower limits would increase the number of Cherenkov photons emitted and therefore, increase the number of events triggering the telescope arrays. Nevertheless, such increase could be compensated by a smaller $\epsilon_{\mathrm{BDT}}$ value when considering completely background free observations. Moreover, the actual field of view of CTA-North telescopes is larger than the viewing cone angle $\alpha$ used to simulate diffuse sources. Consequently, the $\Omega$ factor is expected to be slightly larger than what was computed here in the case of diffuse sources. Finally, we also expect the effective area to significantly increase with the energy of the UHE photons as EASs would be initiated by higher energy bremsstrahlung photons.

In order to calculate the number of preshower events that can be expected in the observation configuration presented in our study, we need an estimation of the UHE photon flux $\phi_{\mathrm{\gamma}}(E)$ from various production models. First of all, we consider the diffuse UHE photon production models such as Super-Heavy Dark Matter (SHDM) decay \cite{shdm06} and GZK photon emission \cite{gelmini08}, as well as the limits put by the Pierre Auger Observatory on the the integral diffuse UHE photon flux \cite{rautenberg19}. The number of expected preshowers is defined as:
\begin{equation}
    N_{\mathrm{preshw}}(E)=\phi_{\mathrm{\gamma-_{\mathrm{diff}}}}(40\mbox{ }\mathrm{EeV})\epsilon_{\mathrm{conv}}\Delta tA_{\mathrm{eff}},
    \label{eqn:N_e}
\end{equation}

where $\epsilon_{\mathrm{conv}}$ is the probability for a UHE photon to produce a preshower in the previously given direction and is equal to 0.67, $\Delta t$ is the observation time and $A_{\mathrm{eff}}$ is the effective area previously calculated. The values taken for the SHDM model, the GZK emission and the Pierre Auger Observatory limits at 40 EeV, expresssed in $\mathrm{km^{-2}yr^{-1}sr^{-1}}$, are $2.7\times10^{-2}$, $4.1\times10^{-4}$ and $3.4\times10^{-4}$, respectively. Such diffuse flux must be multiplied by $4\pi$ when considering point sources of UHE photons. The number of preshowers that can be expected from these scenarios and for a typical observation time of 30 hours is shown in Table \ref{tab:N_E}. 

Focusing our attention on targeted searches for point sources of UHE photons by different experiments, we can provide an estimation of the number of expected preshowers from upper limits put on the UHE photon flux emerging from such sources as well as from possible flares that could potentially "boost" UHE photon emission. While searches for UHE photon point sources were performed by the Pierre Auger Observatory \cite{aab17}, they were limited to the energy range $10^{17.3}<E<10^{18.5}$ eV. However, the extrapolation of the TeV $\gamma$-ray flux of the Galactic Center measured by H.E.S.S. \cite{hess16} to the EeV domain and performed in \cite{aab17} puts an upper limit to 0.034 $\mathrm{km^{-2}yr^{-1}}$ in this energy range. Upper limits reported by the Telescope Array collaboration above 31.6 EeV give an average of 0.0071 $\mathrm{km^{-2}yr^{-1}}$ for the point-source UHE photon flux (see Table 2 of \cite{abbasi20}). Sky maps of upper limits at 95$\%$ C.L. are also displayed in Figure 3 of \cite{abbasi20} and give upper limits up of 0.019 $\mathrm{km^{-2}yr^{-1}}$ near the equator for energies above 31.6 EeV, in the most optimistic case. These values are reported in the second row of Table \ref{tab:N_E_p}. Using Equation \ref{eqn:N_e}, we can estimate the number of preshowers expected from these point-source photon flux upper limits in non-transient mode. Results are reported in the third row of Table \ref{tab:N_E_p}.

\begin{table*}[!ht]
\centering
\begin{tabular}{clll}
 & \multicolumn{1}{c}{AUGER$_{\mathrm{point}}$} & \multicolumn{1}{c}{$\langle \mathrm{TA}_{E> 31.6\:\mathrm{EeV}} \rangle$} & \multicolumn{1}{c}{max(TA$_{E> 31.6\:\mathrm{EeV}}$)}\\ \hline \hline
$\phi_{\mathrm{\gamma-_{\mathrm{p.}}}}(40\mbox{ }\mathrm{EeV})$ [$\mathrm{km^{-2}yr^{-1}}$] & \multicolumn{1}{c}{0.034} & \multicolumn{1}{c}{0.0073} & \multicolumn{1}{c}{0.019} \\ \hline      
$N_{\mathrm{preshw}}$ -- non-transient & \multicolumn{1}{c}{$2.7\times10^{-4}$} & \multicolumn{1}{c}{$5.7\times10^{-5}$} & \multicolumn{1}{c}{$1.5\times10^{-4}$} \\
 -- $R=5$ & \multicolumn{1}{c}{$1.4\times10^{-3}$} & \multicolumn{1}{c}{$2.9\times10^{-4}$} & \multicolumn{1}{c}{$7.6\times10^{-4}$}\\
\ \ \ \ -- $R=652$ & \multicolumn{1}{c}{0.17} & \multicolumn{1}{c}{0.037} & \multicolumn{1}{c}{0.09} \\
\hline \hline
\end{tabular}
\caption{\textit{Point sources of UHE photons} -- Summary of the number of preshowers expected in the case of point sources of UHE photons. The UHE photon flux in the second row is obtained from the upper limits put by the Pierre Auger Observatory on the extrapolation of the H.E.S.S. measurements of the Galactic Center in the energy range $10^{17.3}$ eV - $10^{18.5}$ eV, and from upper limits derived by the Telescope Array, as described in the text. The number of expected preshowers are obtained using Equation \ref{eqn:N_e} for these upper limits and for UHE photons emission boosted by a factor $R$ in the case of GRBs observed by MAGIC and H.E.S.S..}
\label{tab:N_E_p}
\end{table*}

In 2019, the H.E.S.S. telescope array observed the very-high energy emission of the afterglow of gamma-ray burst (GRB) GRB 180720B, several hours after the prompt emission phase \cite{hess19}. The spectrum measured from this observation in the energy range 100 -- 440 GeV is characterized by a flux normalization of $7.52\times10^{-10}\:\mathrm{TeV^{-1}cm^{-2}s^{-1}}$ at 0.154 TeV. In the same year, the MAGIC observatory observed the flaring of GRB 190114C for several hours, one minute after its first detection by other experiments \cite{magic19}. In this observation, the fitting of the spectrum, after correction from the Extragalactic Background Light (EBL) absorption, puts a flux of $9.59\times10^{-8}\:\mathrm{TeV^{-1}cm^{-2}s^{-1}}$ at 0.154 TeV. From these values, we can derive a factor $R$ defined as the ratio of the emission flux in the afterglow or flaring phase to the one of the Galactic Center in non-transient emission mode, at 0.154 TeV, estimated by the H.E.S.S. observatory \cite{hess16} ($1.47\times10^{-10}\:\mathrm{TeV^{-1}cm^{-2}s^{-1}}$\footnote{This value is calculated from a power-law, $dN/dE = \psi(E/\mathrm{TeV})^{-\alpha}$, where $\psi=1.92\times10^{-12}\:\mathrm{TeV^{-1}cm^{-2}s^{-1}}$ is the flux normalization and $\alpha=2.32$ is the spectral index.}). In the case of H.E.S.S. afterglow observation, this ratio is approximately equal to 5 while in the case of MAGIC flare detection, we get $R=652$. Such ratio provides an order of magnitude of the amount by which photon emission can be boosted when a GRB is occuring. Assuming that this boost can be extended to the EeV range, we calculate the number of expected preshower events for both given values of $R$ by simply multiplying the second row of Table \ref{tab:N_E_p} with $R$. Results are shown in the fourth and fifth row of Table \ref{tab:N_E_p}. In these scenarios, the expected number of preshowers is significantly higher than in the case of diffuse emission of UHE photons, reaching up to 0.17 preshower events in 30 hours observation time.

Larger maximum impact distances $R_{\mathrm{max}}$ as well as different observation mode (such as one described in \cite{neronov16}, where telescopes are pointing in different directions just above the horizon) could increase the aperture and therefore, the number of expected preshowers. In fact, a compromise could be found if the pointing direction of each telescope was slightly shifted relative to one another in order to cover a larger strip of the atmosphere above the horizon and to look for a diffuse flux. Letting their field of view overlap would also allow stereoscopic observations.

\section{Discussion and Summary}

In this paper, we investigated the feasibility of observing very inclined air showers produced by the \textit{preshower} effect --- a phenomenon describing the production of an electromagnetic cascade by a UHE photon above the atmosphere --- with the next generation of gamma-ray telescopes developed by the CTA collaboration. The properties of such cascades were studied. Our results showed that in the case of CTA-North site, the conversion probability was higher for high zenith angles and northern directions. The properties of the EASs produced by preshowers, CR background and unconverted UHE photons were also examined. The differences in the $X_{\mathrm{max}}$ distributions and muonic component were noted. The preshower/hadron separation obtained via the analysis of the shape of the images recorded by the IACTs cameras and produced by both CR background, from 10 TeV to 10 EeV, and preshower-induced EASs was proved to be excellent, with signal efficiencies above 90\% (with small background contamination) and above 80\% for background-free observations. Using these results, we calculated the aperture as well as the number of expected preshowers for different UHE photon production models  as well as for upper limits set by the Pierre Auger Observatory and Telescope Array for point and diffuse sources. Although such number was found to be low, the potential of Cherenkov detectors in adopting different observation modes (nearly horizontal direction and/or extended mode) in order to discriminate UHE photon primaries from CR background was underlined. Moreover, at 40 EeV, the required integral diffuse photon flux to obtain $N_{\mathrm{preshw}}=1$ for a 30 hours observation time and for an aperture of 3.42 $\mathrm{km^{2}}$ was calculated to be around 10.1 $\mathrm{km^{-2}year^{-1}sr^{-1}}$. This value is significantly above the prediction given by the UHE photons production models and the upper limits on the integral diffuse UHE photon flux put by the Pierre Auger Observatory mentioned previously. However, in the case of point sources and GRB flares with emission boost of 652, the minimum required flux to obtain $N_{\mathrm{preshw}}=1$ is estimated around 0.2 $\mathrm{km^{-2}year^{-1}}$, which is approximately one order of magnitude higher than upper limits put by the Pierre Auger Observatory and Telescope Array.

We have shown in this paper that the IACT technique could be used to probe physical phenomena not only in the TeV domain, but also in the EeV regime. Although the rate of expected preshower events is quite low, the gamma/hadron separation obtained by adopting the nearly-horizontal observation mode allows for strong filters to be applied in order to identify such events with high degree of confidence. Searches for particles with low expected flux using IACTs have been previously performed, as it is the case for the tau neutrino \cite{magic18} or UHE cosmic rays, as demonstrated in \cite{neronov16}. Moreover, multimessenger alerts obtained from other operating observatories may allow a fast pointing of the telescopes towards cosmic events suspected to be capable of producing UHE photons, such as interactions between UHE cosmic rays possibly produced by AGNs and the cosmic microwave background, or gamma-ray bursts. An alert system would significantly increase the chance probability to observe UHE photons. Such potential is well illustrated by the detection of a 290 TeV neutrino by IceCube and of which the direction is highly correlated with blazar TXS 0506+056 observed by MAGIC and FERMI-LAT. Moreover, a program of observation could be run on catalogs of high energy sources, with observation time significantly higher than the 30h presented in our study.

\section{Acknowledgments}

This research has been supported in part by PLGrid Infrastructure. We warmly thank the staff at ACC Cyfronet AGH-UST, for their always helpful supercomputing support. We also thank K. Bernl\"ohr
for his kind assistance regarding the \textit{sim\_telarray} package.

\section*{References}
\bibliographystyle{elsarticle-num-names} 
\bibliography{bibl}


\end{document}